\documentclass[letterpaper,12pt]{article}

\usepackage{cite}

\usepackage{fancyheadings}
\usepackage[dvips]{graphicx}
\usepackage{graphpap}
\usepackage{ifthen}
\usepackage{enumerate}
\usepackage{amssymb}
\usepackage[mathscr]{eucal}
\usepackage[errorshow]{tracefnt}
\usepackage{fancybox}
\usepackage{amsfonts}
\usepackage{amsmath}
\usepackage{amssymb}
\usepackage{amsthm}
\usepackage{eucal}
\usepackage{eufrak}
\usepackage{multirow}
\usepackage{array}
\usepackage{mathtools}

\makeatletter
\newcommand{\thickhline}{%
    \noalign {\ifnum 0=`}\fi \hrule height 1pt
    \futurelet \reserved@a \@xhline
}
\newcolumntype{'}{@{\hskip\tabcolsep\vrule width 1pt\hskip\tabcolsep}}
\makeatother
\newcolumntype{"}{@{\hskip\tabcolsep\vrule width 1.5pt\hskip\tabcolsep}}
\makeatother

\newcommand{\cA}{\mathcal{A}}
\newcommand{\cB}{\mathcal{B}}
\newcommand{\cC}{\mathcal{C}}
\newcommand{\cD}{\mathcal{D}}

\newcommand{\cG}{\mathcal{G}}
\newcommand{\cH}{\mathcal{H}}

\newcommand{\cM}{\ensuremath{\mathcal{M}}}
\newcommand{\cN}{\ensuremath{\mathcal{N}}}
\newcommand{\cP}{\ensuremath{\mathcal{P}}}
\newcommand{\cQ}{\ensuremath{\mathcal{Q}}}
\newcommand{\cR}{\ensuremath{\mathcal{R}}}
\newcommand{\cS}{\ensuremath{\mathcal{S}}}
\newcommand{\cT}{\ensuremath{\mathcal{T}}}

\newcommand{\cV}{\ensuremath{\mathcal{V}}}

\newcommand\nn{\nonumber}

\newcommand{\uglsat}{U}     

\newcommand{\beq}{\begin{equation}}
\newcommand{\beqn}{\begin{equation*}}
\newcommand{\eeq}{\end{equation}}
\newcommand{\eeqn}{\end{equation*}}
\newcommand{\beqa}{\begin{eqnarray}}
\newcommand{\beqan}{\begin{eqnarray*}}
\newcommand{\eeqa}{\end{eqnarray}}
\newcommand{\eeqan}{\end{eqnarray*}}
\newcommand{\bdm}{\begin{displaymath}}
\newcommand{\edm}{\end{displaymath}}

\newcommand{\la}{\langle}
\newcommand{\ra}{\rangle}

\newtheorem{thm}{Theorem} [section]

\def\Pf{\noindent \textbf{Proof. }}
\def\End{\mathrm{End\ }}

\def\dim{\mathrm{dim\,}}

\def\der'{\mathfrak{der}'\,}
\def\der{\mathfrak{der}\,}
\def\str'{\mathfrak{str}'\,}
\def\str{\mathfrak{str}\,}

\def\g{\gamma}

\def\frake{\mathfrak{e}}

\def\g{\mathfrak{g}}

\def\sl{\mathfrak{sl}}

\def\qed{\hspace{\stretch{1}} $\square$ \\
\noindent}

\newcommand{\al}{\alpha}
\newcommand{\be}{\beta}

\newcommand{\de}{\delta}
\newcommand{\ga}{\gamma}

\newcommand{\Th}{\Theta}

\newcommand{\dlb}{\ensuremath{[\![}}
\newcommand{\drb}{\ensuremath{]\!]}}

\RequirePackage{hyperref}

\numberwithin{equation}{section}

\begin{document}

\vskip-10pt
\hfill {\tt \today}\\

\pagestyle{empty}

\vspace*{2.5cm}

\noindent
\begin{center}
{\LARGE {\sf \textbf{ 
The tensor hierarchy algebra}}}\\
\vspace{.3cm}

\renewcommand{\thefootnote}{\fnsymbol{footnote}}

\vskip 1truecm
\noindent
{\large {\sf \textbf{Jakob Palmkvist}
}}\\
\vskip 1truecm
        {\it 
        {Institut des Hautes Etudes Scientifiques\\ 35, Route de Chartres, FR-91440 Bures-sur-Yvette, 
France}\\[3mm]}
        {\tt palmkvist@ihes.fr} \\
\end{center}

\vskip 1cm

\centerline{\sf \textbf{
Abstract}}
\vskip .2cm

\noindent
We introduce an infinite-dimensional Lie superalgebra which is an extension of the U-duality Lie algebra 
of maximal supergravity in $D$ dimensions, for $3 \leq D \leq 7$.
The level decomposition with respect to the U-duality Lie algebra gives exactly the tensor hierarchy of representations 
that arises in gauge deformations of the theory described by an embedding tensor, for all positive levels $p$. 
We prove that these representations are always contained in those coming from the associated Borcherds-Kac-Moody superalgebra,
and we explain why some of the latter representations are not included in the tensor hierarchy.
The most remarkable feature of our Lie superalgebra is that it does not admit a triangular decomposition like a (Borcherds-)Kac-Moody (super)algebra. Instead the Hodge duality relations between level $p$ and $D-2-p$ extend to negative $p$, relating the representations at the first two negative levels to the supersymmetry and closure constraints of the embedding tensor. 

\newpage

\pagestyle{plain}

\tableofcontents

\section{Introduction}

Gauged supergravity has generated much interest
since it can be used to describe compactifications of string theory, or possibly even M-theory, to lower dimensions.
Already thirty years ago
a four-dimensional theory with an SO(8) gauge group was constructed \cite{deWit:1981eq}, and recently it
has been shown that there exists a continuous one-parameter
family of such theories \cite{Dall'Agata:2012bb}
(see also \cite{Borghese:2012qm,Borghese:2012zs,Borghese:2013yp,Borghese:2013dja,Dall'Agata:2012sx,deWit:2013ija,Godazgar:2013nma}).
This development made use of the embedding tensor approach, which has 
proven to be a powerful tool
in the systematic study of gauged supergravity
in various dimensions \cite{Nicolai:2000sc,Nicolai:2001sv,deWit:2002vt,deWit:2003hr,deWit:2004nw,Samtleben:2005bp,Weidner:2006rp,
deWit:2007mt,Bergshoeff:2007ef}.

We will in this paper consider gauged supergravity theories with maximal supersymmetry in $D$ spacetime dimensions, where $3 \leq D \leq 7$. 
These theories are deformations
of the corresponding ungauged theories, which in turn can be
obtained from eleven-dimensional supergravity \cite{Cremmer:1978km}
by dimensional reduction. The global symmetry (or U-duality) that arises
in the reduction is then broken by the deformation, which promotes only a part of it to a local symmetry.
The embedding tensor describes how the corresponding gauge group ${\rm G}_0$ is embedded into the global symmetry group ${\rm G}$.
Treating it as a spurionic object makes it possible to formulate the theory in a ${\rm G}$-covariant way,
although the ${\rm G}$-invariance is broken down to ${\rm G}_0$ as soon as the values of the embedding tensor are fixed, and thus the gauge group specified.

The G-covariant formulation
requires a hierarchy of $p$-form fields, transforming in a sequence of representations of the global symmetry group ${\rm G}$, or of its (complexified)
Lie algebra $\g=\mathfrak{e}_{11-D}$,
which is simple and finite-dimensional for $3 \leq D \leq 7$ \cite{Cremmer:1978ds,Julia:1982gx,Cremmer:1997ct}.
This sequence can be considered as infinite, although the $p$-form fields themselves collapse to zero whenever the number $p$ of antisymmetric spacetime indices exceeds the
dimension $D$.

Remarkably, 
the tensor hierarchy can be derived from the Lie algebra $\g$ by extending it to either a Borcherds-Kac-Moody (BKM) superalgebra
\cite{HenryLabordere:2002dk,Henneaux:2010ys}, or to the Kac-Moody algebra
$\mathfrak{e}_{11}$ \cite{Riccioni:2007au,Bergshoeff:2007qi,Bergshoeff:2007vb,Riccioni:2007ni,Riccioni:2009hi,Riccioni:2009xr,Henneaux:2010ys}, both infinite-dimensional. Up to the spacetime limit $p=D$,
the representations are then found at the positive levels
in the level decomposition with respect to $\g$ or $\g \oplus \mathfrak{sl}(D)$, respectively
(in the $\mathfrak{e}_{11}$ case restricted to antisymmetric $\mathfrak{sl}(D)$ representations)
This `empirical' fact that was explained in \cite{Palmkvist:2011vz}.
The same representations appear in other contexts as well, where it is important to know how the sequence continues beyond the spacetime limit $p=D$,
for example in exceptional generalized geometry \cite{Berman:2012uy,Cederwall:2013naa,Cederwall:2013oaa}.  
It is also important
from a superspace point of view \cite{Greitz:2011vh,Greitz:2011da,Greitz:2012vp},
since $p$-forms in superspace can be non-zero even if $p > D$.
However, it is less clear how the tensor hierarchy is related to the BKM superalgebra beyond the spacetime limit. A deeper understanding of its
algebraic structure is needed, something that we aim to reach in this paper.

The representations at levels between (and including) $p=-D+2$ and $p=D-2$ in the BKM superalgebra are `reflected' in two different points at the $p$-axis: $p=0$ and $p=(D-2)/2$. With this we mean that the representation at level $p$ is the conjugate of both the representation at level $-p$ and the one at level $D-2-p$. The latter symmetry has a physical meaning in the sense that it corresponds to the Hodge duality between $p$-forms and $(D-2-p)$-forms. The
symmetry around zero has 
of course no such meaning in the interpretation where 
only the positive levels correspond to $p$-forms, but on the other
hand it is related to a local symmetry, corresponding to the maximal compact subgroup of G, that makes this interpretation possible as a gauge choice.
When going to levels $p \geq D-1$ the symmetry around $(D-2)/2$ gets broken,
whereas the symmetry around
zero remains, being a fundamental property of the algebra.

In this paper we will investigate the possibility of keeping the symmetry around $(D-2)/2$ instead of the one around zero.
Our investigations result in an algebra where the embedding tensor can be interpreted as the components of an element $\Theta$ at level $-1$
that squares to zero, $\{ \Theta, \Theta \}=0$, and the intertwiners that define the tensor hierarchy \cite{deWit:2008ta,deWit:2008gc} are then simply the components of the adjoint action of
$\Theta$. As a consequence, the representations at {\it all}
positive levels in this algebra {\it exactly} coincides with those coming from the tensor hierarchy.
Furthermore, we show that they are always contained in the 
representations coming from the BKM superalgebra.
There are representations in the level decomposition of the BKM superalgebra that the tensor hierarchy misses (not only for $D=3$) but never the other way around.

Since the algebra that we introduce gives exactly the tensor hierarchy, we call it a tensor hierarchy algebra. In our construction it depends on 
an arbitrary simple finite-dimensional complex Lie algebra $\g$, a certain extension of $\g$ to an affine Kac-Moody algebra
$\g_{\rm KM}$, and an integer $D \geq 3$. We are particularly interested in the cases $\g=\mathfrak{e}_{11-D}$
corresponding to maximal supergravity in $D$ dimensions, and many of our results can be obtained
explicitly for each of these cases,
but we make an effort to be as general as possible.

This paper is organized as follows. 
We start with a very brief review of the embedding tensor formalism and the tensor hierarchy in section \ref{gaugra-review}, to give
a physical motivation to the mathematical results in the rest of the paper.
Section \ref{grad-sec} then provides the necessary tools for the construction and study of the
Lie superalgebras that we consider in the following two sections: The BKM superalgebra in section \ref{borcherds-sec},
and the tensor hierarchy algebra in section \ref{tha-sec}. 
We end with some concluding remarks in section \ref{con-sec}. A few proofs are relegated to appendix \ref{superapp},
and some useful formulas for the cases $\g=\mathfrak{e}_{11-D}$ are collected in appendix \ref{explapp}.

\section{The tensor hierarchy in gauged supergravity} \label{gaugra-review}

In this section we will very briefly review how the tensor hierarchy arises in the embedding tensor approach to maximal gauged supergravity.
We will follow \cite{deWit:2008ta} and refer to this paper (and the references therein) for more information. The section is a
slightly shortened and rewritten version of section 2 in \cite{Palmkvist:2011vz}.

We start with the vector field in $D$-dimensional maximal supergravity,
which transforms in an irreducible representation ${\bf r}_1$ of the global symmetry group ${\rm G}$, or of the corresponding Lie algebra $\g$.
We write the vector field as $A_\mu{}^\cM$, where the indices $\cM$ are associated to 
${\bf r}_1$ and $\mu=1,2,\ldots,D$ are the spacetime indices.
By the gauge deformation, a subgroup $\rm{G}_0$ of the global symmetry group $\rm{G}$ is promoted to a local symmetry group, 
with the vector field
as the gauge field. Accordingly, the Lie algebra of the gauge group is spanned by elements $X_\cM$,
with an ${\bf r}_1$ index downstairs.
However, the elements $X_\cM$ need not be linearly independent, so the dimension of the gauge group $\rm{G}_0$ can be smaller than the dimension of ${\bf r}_1$.

We let
$t_\alpha$ be a basis of $\g$, with an adjoint index $\alpha$ 
that we can raise with the inverse of the Killing form.
Since the gauge group 
$\rm{G}_0$ is a subgroup of $\rm{G}$, the elements $X_\cM$ must be linear combinations of $t_\alpha$ and can be written
$X_\cM = \Theta_\cM{}^\alpha t_\alpha$.
The coefficients of the linear combinations form a tensor $\Theta_\cM{}^\alpha$ which is called the {\it embedding tensor} since it describes how $\rm{G}_0$ is embedded into $\rm{G}$.

It follows from the index structure of the embedding tensor that
it transforms in the tensor product of $\overline{\bf r}_1$, the conjugate of ${\bf r}_1$, and the adjoint
of $\g$. This tensor product decomposes into a direct sum of 
irreducible representations, but supersymmetry restricts the embedding tensor to only one or two of them. We will refer to this
restriction as the {\it supersymmetry constraint}. In addition, gauge invariance of the embedding tensor leads to the condition
\begin{align} \label{closconstr}
0&=\de_\cM(\Th_\cN{}^\al)
=-\Th_\cM{}^\be (t_\be)_{\cN}{}^\cP \Th_\cP{}^\al
+\Th_\cM{}^\be f_{\be\ga}{}^\al\Th_\cN{}^\ga,
\end{align}
which contracted with $t_\alpha$ turns into
\begin{align}
[X_\cM,X_\cN]=(X_\cM)_\cN{}^\cP X_\cP, \label{strukturkonstantekv}
\end{align}
and thus ensures that the gauge group closes under the commutator. It is therefore called the {\it closure constraint}.
In (\ref{strukturkonstantekv}) the components $(X_\cM)_\cN{}^\cP=\Theta_\cM{}^\al (t_\al)_\cN{}^\cP$ 
look like structure constants for the gauge group, but because of the possible linear dependence in the set of elements $X_\cM$, the
components $(X_\cM)_\cN{}^\cP$ are in general not antisymmetric in the lower indices. Only when we contract $(X_\cM)_\cN{}^\cP$ with another 
$X_\cP$ the symmetric part vanishes.

Before proceeding we want to make the reader aware of our conventions, which result in a different
sign of the first term on the right hand side of (\ref{closconstr}), and of the only term on the right hand side of (\ref{strukturkonstantekv}),
compared to \cite{deWit:2008ta} (and many other references).
To obtain agreement with \cite{deWit:2008ta}, $(t_\alpha)_\cM{}^\cN$ should be replaced by $-(t_\alpha)_\cM{}^\cN$ everywhere.
(Unfortunately this sign was missing in \cite{Palmkvist:2011vz}.)

When we gauge the theory we replace the partial derivatives with covariant ones,
\begin{align}
\partial_\mu \to D_\mu = \partial_\mu-gA_\mu{}^\cM X_\cM,
\end{align}
where $g$ is a coupling constant.
Consistency of the theory then requires a hierarchy of $p$-form fields 
$A_{\mu_1\cdots\mu_p}{}^{\cM_1\cdots\cM_p}$
which besides the $p$ antisymmetric spacetime indices also carry $p$ indices associated to ${\bf r}_1$,
and thus each transforms in a subrepresentation ${\bf r}_p$ of
the tensor product $({\bf r}_1)^p$.
Each $(p+1)$-form field $A_{\mu_1\cdots\mu_{p+1}}{}^{\cM_1\cdots\cM_{p+1}}$ 
arises in contributions to the gauge transformation
of the $p$-form field strength via an intertwiner
$Y^{\cM_1\cdots\cM_{p}}{}_{\cN_1\cdots\cN_{p+1}}$
where the upper and lower set of indices correspond to the representations ${\bf r}_{p}$ and ${\bf r}_{p+1}$, respectively.
These intertwiners 
are defined recursively by 
the formula
\begin{align}
Y^{\cM_1\cdots\cM_p}{}_{\cN_1\cdots\cN_{p+1}}
&=(X_{\cN_1})_{\cN_2\cdots\cN_{p+1}}{}^{\la\cM_1\cM_2\cdots\cM_p\ra}-\de_{\cN_1}{}^{\la\cM_1}Y^{\cM_2\cdots\cM_p\ra}{}_{\cN_2\cdots\cN_{p+1}}
\label{y-relation}
\end{align}
and the initial condition
\begin{align}
Y^\cP{}_{\cM\cN} = (X_\cM)_\cN{}^\cP + (X_\cN)_\cM{}^\cP, \label{lilla1}
\end{align} 
where the angle brackets in (\ref{y-relation}) denote projection on ${\bf r}_p$. The lower indices of
the intertwiner
$Y^{\cM_1\cdots\cM_p}{}_{\cN_1\cdots\cN_{p+1}}$ then define ${\bf r}_{p+1} \subseteq {\bf r}_p \times {\bf r}_1$ so that, by definition,
\begin{align}
Y^{\cM_1\cdots\cM_p}{}_{\cN_1\cdots\cN_{p+1}}=Y^{\cM_1\cdots\cM_p}{}_{\la\cN_1\cdots\cN_{p+1}\ra}.
\end{align}
(Obviously we also have
$Y^{\cM_1\cdots\cM_p}{}_{\cN_1\cdots\cN_{p+1}}=Y^{\la \cM_1\cdots\cM_p \ra}{}_{\cN_1\cdots\cN_{p+1}}$.)
The formulas (\ref{y-relation}) and (\ref{lilla1}) define a sequence of 
representations ${\bf r}_p$ of $\g$ for all positive integers $p$, also for $p>D$ since no spacetime indices enter.
The only input is $\g$ itself, ${\bf r}_1$ and the supersymmetry constraint (which is needed to determine ${\bf r}_2$).
Below we list $\mathfrak{g}$ and ${\bf r}_p$ for $3 \leq D \leq 7$ and $1 \leq p \leq D$ \cite{deWit:2008ta,deWit:2008gc,deWit:2009zv}.
\setlength{\arraycolsep}{9.95pt}
\\
{\renewcommand{\arraystretch}{1.5}
\begin{align*}
\begin{array}{|c|c|c|c|c|c|c|c|c|}
\hline
D&\g&{\bf r}_1&{\bf r}_2&{\bf r}_3&{\bf r}_4&{\bf r}_5&{\bf r}_6&{\bf r}_7\\
\hline
\multirow{2}{*}{7}&\multirow{2}{*}{$\mathfrak{a}_4$}&\multirow{2}{*}{$\overline{\bf 10}$}&\multirow{2}{*}{${\bf 5}$}&\multirow{2}{*}{$\overline{\bf 5}$}&\multirow{2}{*}{${\bf 10}$}&\multirow{2}{*}{${\bf 24}$}&\multirow{2}{*}{$\overline{\bf 15}+{\bf 40}$}&{\bf 5}+\overline{\bf 45}\\
&&&&&&&&+\,{\bf 70}\\\hline
\multirow{2}{*}{6}&\multirow{2}{*}{$\mathfrak{d}_5$} &\multirow{2}{*}{${\bf 16}_c $}&\multirow{2}{*}{$ {\bf 10} $}&\multirow{2}{*}{${\bf 16}_s$}&\multirow{2}{*}{${\bf 45}$}&\multirow{2}{*}{${\bf 144}_s$}&{\bf 10}+{\bf 126}_s&\\
&&&&&&&+\,{\bf 320}&\\\hline
\multirow{2}{*}{$5$}&\multirow{2}{*}{$\mathfrak{e}_6$}  &\multirow{2}{*}{$\overline{\bf 27}  $}&\multirow{2}{*}{$ {\bf 27} $}&\multirow{2}{*}{${\bf 78} $}&\multirow{2}{*}{${\bf 351}$}& {\bf 1728}&&\\
&&&&&&+\,{\bf 27}&&\\\hline
\multirow{2}{*}{$4$}&\multirow{2}{*}{$\mathfrak{e}_7$} &\multirow{2}{*}{${\bf 56}$} &\multirow{2}{*}{$ {\bf 133} $}&\multirow{2}{*}{${\bf 912}$}&{\bf 8645}&&&\\
&&&&&+\,{\bf 133}&&&\\\hline
\multirow{2}{*}{$3$}&\multirow{2}{*}{$\mathfrak{e}_8$} &\multirow{2}{*}{${\bf 248}  $}&\multirow{2}{*}{$ {\bf 3875}$}&{\bf 147250}&&&&\\
&&&&+\,{\bf 3875}&&&&\\
\hline
\end{array}
\end{align*}
}
Although no spacetime indices enter in the formula (\ref{y-relation}), the table shows that the representations know about spacetime. 
The duality between $p$-forms and $(D-2-p)$-forms 
is related to conjugation of the corresponding representations,
${\overline{\bf r}_p} = {\bf r}_{D-2-p}$.
Furthermore, ${\bf r}_{D-2}$ is always the adjoint {\bf adj} of $\g$, and the last two representations in each row are related to the constraints of the embedding tensor: ${\overline{\bf r}_{D-1}}$ is contained in the subrepresentation 
of ${\bf adj} \times {\overline{\bf r}_1}$ in which the embedding tensor must transform according to the supersymmetry constraint,
and ${\overline{\bf r}_{D}}$ is the representation in which the closure constraint transforms.

For each $D$ the sequence of representations ${\bf r}_p$ can be obtained from a graded Lie superalgebra that we call a tensor hierarchy algebra,
and denote by $\cT$. It can in turn be constructed from a BKM superalgebra $\cB$, 
which also directly defines a sequence of representations ${\bf s}_p$ such that ${\bf r}_p\subseteq{\bf s}_p$ (and thus ${\bf r}_p={\bf s}_p$
if ${\bf s}_p$ is irreducible). However, it is not always true that ${\bf r}_p={\bf s}_p$, as can be seen for $D=3$ by comparing the table above with the one in appendix \ref{app-b2} (where also the Dynkin labels of the representations are written out).
In order to explain this in detail, we need to switch to a more mathematical point of view.

\section{Graded Lie superalgebras} \label{grad-sec}

Any Lie superalgebra is by definition a $\mathbb{Z}_2$-graded algebra, but 
when we talk about graded Lie superalgebras in this paper we assume that
the $\mathbb{Z}_2$-grading is refined into a 
$\mathbb{Z}$-grading, consistent with the $\mathbb{Z}_2$-grading. 
For each of the two gradings there is a decomposition of the algebra into a direct sum of subspaces, labelled by the elements in the corresponding set. Whenever
$\mathcal{G}$ is a graded Lie superalgebra, we will write 
these subspaces as $\mathcal{G}_{(p)}$ for $p \in \mathbb{Z}_2$, and
$\mathcal{G}_{p}$ for $p \in \mathbb{Z}$. 
Thus 
\begin{align}
\mathcal{G} = \mathcal{G}_{(0)} \oplus \mathcal{G}_{(1)} = 
\cdots \oplus \mathcal{G}_{-1} \oplus \mathcal{G}_{0} \oplus \mathcal{G}_{1} \oplus \cdots,
\end{align}
and consistency means that $\mathcal{G}_{p} \subseteq \mathcal{G}_{(q)}$ whenever $p \equiv q$ (mod 2).
An element that belongs to one of the subspaces of a grading is said to be homogeneous, and for the $\mathbb{Z}_2$-grading we then write
$|x|=p$ if $x \in \mathcal{G}_{(p)}$.

We write the supercommutator of any two elements $x$ and $y$ as
$\dlb x,y\drb$. 
If $x$ and $y$ are homogeneous elements
we may replace the supercommutator by the commutator $[x,y]$ or the anticommutator $\{ x,y\}$. It then satisfies the
(anti)symmetry
\begin{align}
\dlb x , y \drb &= -(-1)^{|x||y|} \dlb y , x \drb
\end{align}
and the Jacobi identity
\begin{align}
\dlb x , \dlb y, z \drb \drb &= \dlb \dlb x, y \drb, z \drb +(-1)^{|x||y|} \dlb y, \dlb x,z\drb\drb.
\end{align}
In a graded Lie superalgebra the
supercommutator respects not only the $\mathbb{Z}_2$-grading,
$\dlb \mathcal{G}_{(p)}, \mathcal{G}_{(q)} \drb = \mathcal{G}_{(p+q)} \ \text{(mod 2)},$
but also the $\mathbb{Z}$-grading,
$\dlb \mathcal{G}_{p}, \mathcal{G}_{q} \drb = \mathcal{G}_{p+q}$.
In particular the adjoint action induces a representation of the subalgebra $\mathcal{G}_0$ (which is an ordinary Lie algebra)
on each subspace $\mathcal{G}_p$.

We will let $\mathcal{G}_{\pm}$ denote the direct sum of all subspaces $\mathcal{G}_{\pm k}$ with $k > 0$.
Furthermore, it will be convenient to denote by
$\mathcal{G}_{0\pm}$ the direct sum of all subspaces $\mathcal{G}_{\pm k}$ with $k \geq 0$, and to let
$\mathcal{G}_{2+}$ denote the direct sum of all subspaces
$\mathcal{G}_{k}$ with $k \geq 2 $.

\subsection{The universal graded Lie superalgebra}

For any vector space $U_1$,
there is an associated graded Lie superalgebra $\mathcal{U}(U_1)$,
analogous to the
universal graded Lie algebra introduced by Kantor \cite{Kantor-graded}, and thus we may call $\mathcal{U}(U_1)$
the universal graded Lie superalgebra of $U_1$.
Except for a slight change in the definition of the supercommutator, and a sign change in the $\mathbb{Z}$-grading, the construction below
was given in \cite{Palmkvist:2009qq} (in the context of three-algebras).

The subspaces $\mathcal{U}(U_1)_{k}$ for $k \leq 1$
are defined recursively, starting with 
$\mathcal{U}(U_1)_{1}=U_1$, which
enables us to write $\mathcal{U}(U_1)=U$.
For each $k\geq0$, the subspace $\uglsat_{-k}$ is then defined as the vector space of all linear maps
$U_1 \to \uglsat_{-k+1}$.
Thus $\uglsat_0=\End{U_1}$, and $\uglsat_{-1}$ consists of linear maps $U_1 \to \End{U_1}$.
The supercommutator on the subalgebra $U_{0-}$
is defined recursively by
\begin{align}\label{recdefsuperbracket}
\dlb x,y \drb = (\text{ad}^\ast\, y) \circ x - (-1)^{|x||y|} (\text{ad}^\ast\, x) \circ y,
\end{align}
where $\text{ad}^\ast$ denotes adjoint action from the right, $(\text{ad}^\ast\,x)(y)=\dlb y, x \drb$.
(It may seem more natural to employ the usual adjoint action from the left, and this is how the associated graded Lie superalgebra was defined in 
\cite{Palmkvist:2009qq}, but the above choice will turn out to be more convenient for our purposes here.)
The Jacobi {identity} on $U_{0-}$ can then be shown to hold by induction. 
The subspaces $U_k$ for $k\geq 2$ and the supercommutator on the subalgebra $U_+$
are defined such that $U_+$ is the free Lie superalgebra generated by $U_1$.
The supercommutator 
$\dlb x,\,y \drb $ for 
$x \in U_+$ and 
$y \in U_{0-}$ is defined as $y(x)$ if $x \in U_1$. It can then be extended to any
$x \in U_+$
by the Jacobi identity.

Now let $V_{-1}$ be a subspace of $U_{-1}$, and let $\mathcal{V}(U_1,V_{-1})$ be the subalgebra of $U$ generated by $U_1$ and $V_{-1}$.
We write $\mathcal{V}(U_1,V_{-1})=V$ and thus $U_1=V_1$.
An ideal $\cH$ of $\cG$ is graded if it is the direct sum of all subspaces $\cH_p=\cH \cap \cG_p$.
Let $\mathcal{D}_+$ be the direct sum of all graded ideals of $V$
and set
\begin{align}
\mathcal{V}\,'(U_1,V_{-1})=V/\mathcal{D}_+.
\end{align}
The plus sign indicates that $\cD_+$ is automatically contained in $V_+$.

The BKM superalgebra $\cB$ and the tensor hierarchy algebra $\cT$ that we will consider next are both special cases 
of $\mathcal{V}\,'(U_1,V_{-1})$ with the same vector space $U_1$, but with different subspaces $V_{-1} \subseteq U_{-1}$.

\section{The Borcherds-Kac-Moody superalgebra} \label{borcherds-sec}

\subsection{Chevalley-Serre construction} \label{cs-constr}

The BKM superalgebra $\cB$ is an infinite-dimensional extension of a simple finite-dimensional Lie algebra $\g$,
with rank $r$ and Cartan matrix $A_{ij}$, where $i,j=1,2,\ldots,r$.
We then extend $A_{ij}$ by an extra row and column to a matrix $A_{IJ}$, which will be the Cartan matrix of $\cB$, where $I,J=0,1,\ldots,r$.

We require that the diagonal entry $A_{00}$ is equal to zero and that the off-diagonal entries 
are non-positive integers with $A_{0i}=A_{i0}$ such that the determinant is given by
\begin{align} \label{detvillkor}
\det{A_{IJ}} = -\frac{D-1}{D-2} \det{A_{ij}}
\end{align}
for some integer $D \geq 3$ (the reason for this is given by Theorem \ref{affthm} below).

Given its Cartan matrix $A_{IJ}$, the BKM superalgebra $\mathcal{B}$ is
the Lie superalgebra generated by $2(r+1)$ elements $e_i,f_i \in \mathcal{B}_{(0)}$ and $e_0,f_0 \in \mathcal{B}_{(1)}$ modulo the {\it Chevalley relations}
\begin{align} \label{chev-rel0}
[ h_I,e_J ]&=A_{IJ}e_J, & [ h_I,f_J ]&=-A_{IJ}f_J, & \dlb e_I,f_J \drb &=\delta_{IJ}h_J,
\end{align}
and the {\it Serre relations}
\begin{align}
\{e_0,e_0\}=\{f_0,f_0\}=
(\text{ad } e_i)^{1-A_{iJ}} (e_J) = (\text{ad } f_i)^{1-A_{iJ}} (f_J) = 0. \label{serre-rel}
\end{align}
where $h_I = \dlb e_I,f_I\drb \in \mathcal{B}_{(0)}$. It follows from the Chevalley relations (\ref{chev-rel0}) that the Cartan subalgebra,
spanned by the $h_I$, is abelian, $\dlb h_I, h_J \drb=0$.

We stress that $\cB$ is only a very special case of a BKM superalgebra, which in general can have more than one odd simple root (corresponding to $e_0$ and $f_0$ here), and other possible values of the entries in the Cartan matrix. Accordingly, the Chevalley-Serre relations of a general BKM superalgebra are more involved than (\ref{chev-rel0})--(\ref{serre-rel}). We also mention that general BKM superalgebras (of finite rank) in turn are special cases of the contragredient Lie superalgebras introduced in
\cite{Kac77A,Kac77B}. For more information about BKM superalgebras, see \cite{Ray}.

The Dynkin diagram of $\g$ can be extended to a Dynkin diagram of $\mathcal{B}$ by adding an extra node, corresponding to the row and column that we add to $A_{ij}$. We paint it black in order to distinguish it from the other nodes, which we let be white. (This `painting' of nodes follows \cite{HenryLabordere:2002dk}. 
Considering $\cB$ as a contragredient Lie superalgebra, with the conventions of
\cite{Kac77A,Kac77B}, the black node would instead be drawn $\otimes$, and called `gray'. See also \cite{Kleinschmidt:2013em}.)
The off-diagonal entries $A_{0i}=A_{i0}$ of the Cartan matrix can be encoded in the Dynkin diagram by letting $|A_{i0}|$ be the number of lines between node $i$ and node 0 (the black one).
This way of extending the Dynkin diagram is illustrated for $\g=\mathfrak{e}_{11-D}$ below.

\noindent
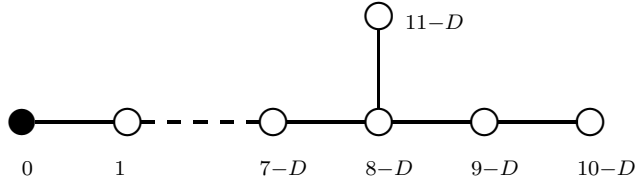
\begin{picture}(450,75)(5,-10)
\put(115,-10){${\scriptstyle{0}}$}
\put(150,-10){${\scriptstyle{1}}$}
\put(205,-10){${\scriptstyle{7-D}}$}
\put(245,-10){${\scriptstyle{8-D}}$}
\put(285,-10){${\scriptstyle{9-D}}$}
\put(325,-10){${\scriptstyle{10-D}}$}
\put(260,45){${\scriptstyle{11-D}}$}
\thicklines
\multiput(210,10)(40,0){4}{\circle{10}}
\multiput(215,10)(40,0){3}{\line(1,0){30}}
\put(155,10){\circle{10}}
\put(115,10){\circle*{10}}
\put(120,10){\line(1,0){30}}
\multiput(160,10)(10,0){5}{\line(1,0){5}}
\put(250,50){\circle{10}} \put(250,15){\line(0,1){30}}
\end{picture} 
\\\\
\noindent
If we replace the black node with a white one, and consider it as the first one in a sequence of $(D-2)$ nodes that we add to the Dynkin diagram of $\g$, each of them connected to the next by a single line, then we obtain the Dynkin diagram of a Kac-Moody algebra $\mathfrak{g}_{\rm KM}$. 
The following proposition about $\mathfrak{g}_{\rm KM}$ is the reason for requiring 
(\ref{detvillkor}) above.

\begin{thm} \label{affthm}
For any integer $D \geq 3$, the Kac-Moody algebra $\mathfrak{g}_{\rm KM}$ is affine if and only if $(\ref{detvillkor})$ holds.
\end{thm}

\Pf
Let $d_k$ $(k \geq 0)$ be the determinant of the Cartan matrix obtained by adding $k$ nodes to the Dynkin diagram of $\g$.
It is easy to verify the recursion formula
\begin{align} \label{diffekv}
d_{k+2}=2 d_{k+1}- d_k
\end{align}
and show that $d_{D-2}=0$ if and only if (\ref{detvillkor}) is satisfied. 
The condition that the determinant of a Cartan matrix be zero is
necessary for it to be affine, but in general not sufficient.
To show that it is actually sufficient in this case, we use the fact that a symmetrizable
Cartan matrix $A$, obtained by adding a node to a Dynkin diagram of a finite Kac-Moody algebra, is either finite (if $\det A > 0$), affine (if $\det A = 0$) or Lorentzian (if $\det A < 0$) \cite{Ruuska} (see also \cite{Gaberdiel:2002db}).
The Cartan matrices that we consider are symmetrizable since we start with a Cartan matrix which is finite (and thus automatically symmetrizable \cite{Kac})
and then extend it symmetrically. 
If we now assume that $d_{D-2}=0$, then it follows from (\ref{diffekv}) that $d_k>0$ for $0 \leq k \leq D-3$, since we know that $d_0 > 0$.
One can then show by induction that the Kac-Moody algebra obtained by deleting the last node from the Dynkin diagram of $\g_{\rm KM}$ is finite, and it
follows that $\g_{\rm KM}$ is affine.
\qed

\noindent
There is always at least one possibility of adding a black node to the Dynkin diagram of $\g$ such that (\ref{detvillkor}) holds, giving as $\g_{\rm KM}$
the affine extension of $\g$, with $D=3$. But for example if $\g=\mathfrak{e}_6$ or $\g=\mathfrak{e}_7$ we can also let $\g_{\rm KM}$ be $\frake_9$, the affine extension of $\mathfrak{e}_8$ (with $D=4$ and $D=5$, respectively) instead of the affine extension of $\g$ itself.

We consider $\cB$ as a graded Lie superalgebra with the $\mathbb{Z}$-grading such that $e_0 \in \cB_1$ and $f_0 \in \cB_{-1}$, whereas $e_i,f_i \in \cB_0$ (and thus $h_I \in \cB_0$).
We let ${\bf s}_{p}$ be the representation of $\g$ on $\cB_p$ given by the adjoint action,
\begin{align}
{\bf s}_p : \quad \g \to \End\cB_p,\quad {\bf s}_p(x)(y)=\dlb x,y \drb.
\end{align}
It follows from the Chevalley relations 
(\ref{chev-rel0}) that
${\bf s}_{p}$ and ${\bf s}_{-p}$ are conjugate to each other, ${\bf s}_{-p}=\bar{\bf s}_p$.
When $\g$ is the global symmetry algebra of maximal supergravity in $D$ dimensions,
${\bf s}_{1}$ is the representation in which the vector field transforms, that is, ${\bf r}_1$ in the tensor hierarchy.
Thus we can use the ${\bf r}_1$ indices introduced in section \ref{gaugra-review}, and let $E_\cM$ and $F^\cM$ be bases of $\mathcal{B}_{1}$ and $\mathcal{B}_{-1}$, respectively, where
$\cM=1,2,\ldots,\text{dim}\,{\bf r}_1$.

As shown in \cite{Palmkvist:2012nc}, the subalgebra $\cB_0$ is the direct sum of $\g$ and a one-dimensional Lie algebra spanned by an element $h$ in the 
Cartan subalgebra. Furthermore, the commutation relations of the basis elements $E_\cM$ and $F^\cM$ (normalized appropriately) with each other and with
$t^\al$ and $h$ are
\begin{align}
\{E_\cM,\,F^\cN\}&=(t_\alpha)_\cM{}^\cN t^\alpha + \frac1{D-2}\de_\cM{}^\cN h, &
[t^\alpha,t^\beta]&=f^{\alpha\beta}{}_\ga t^\ga, & [t^\alpha,h]&=0,\nn
\end{align}
\begin{align}
[t^\alpha,E_\cM]&=(t^\alpha)_\cM{}^\cN E_\cN, & [h,E_\cM]&=-(D-1)E_\cM,\nn\\
[t^\alpha,F^\cN]&=-(t^\alpha)_\cM{}^\cN F^\cM, & [h,F^\cN]&=(D-1)F^\cN. \label{borcherds-comm-rel}
\end{align}

\newpage

\subsection{Construction from the universal graded Lie superalgebra} \label{constrfruglsa}

As an intermediate step in the construction of $\mathcal{B}$ we can first construct a graded Lie superalgebra $\mathcal{A}$ 
with the same generators, grading and Chevalley relations as $\mathcal{B}$, but with
the Serre relations restricted to
\begin{align}
(\text{ad } e_i)^{1-A_{iJ}} (e_J) = (\text{ad } f_i)^{1-A_{iJ}} (f_J) = 0. \label{serre-rel-restr}
\end{align}
For $J=j=1,2,\ldots,r$ these relations are the Serre relations for the $\g$ subalgebra, while for $J=0$ they
fix the subspaces $\cA_{\pm 1}$.

Let $\mathcal{C}_+$ and $\mathcal{C}_-$ be the ideals of $\mathcal{A}$ generated by $\{ e_0,e_0 \}$ and $\{ f_0,f_0 \}$, 
respectively,
the elements which are set to zero in the remaining Serre relations for $\cB$ (\ref{serre-rel}), so that
$\mathcal{B}=\mathcal{A}/(\mathcal{C}_+ \oplus \mathcal{C}_-)$. 
It follows from the relations
\begin{align}
[f_I,\{e_0,e_0\}]=[e_I,\{f_0,f_0\}]=0
\end{align}
in $\cA$ 
that $\mathcal{C}_\pm \subset \mathcal{A}_{\pm}$, and thus $\cA_{\pm 1}=\cB_{\pm 1}$.
It is easy to see that $\{ \cA_{1},\cA_{-1} \}=\cA_0$, so
$\cA$ is generated by $\cA_{\pm 1}$, and one can furthermore show
that the subalgebras $\cA_{\pm}$ are freely generated by $\cA_{\pm 1}$.

For a Kac-Moody algebra, the Gabber-Kac theorem  \cite{Gabber-Kac,Kac} states that the Serre relations
generate the maximal ideal of the `auxiliary algebra', generated by $e_I$ and $f_I$ modulo only the Chevalley relations,
that intersects the Cartan subalgebra trivially. 
This is true also for a Borcherds superalgebra \cite{Ray95,Ray}, and implies in particular
that $\cC_+$ is the maximal graded ideal of $\cA$ contained in $\cA_{2+}$, and that $\cC_-$ is the maximal ideal of $\cA$ contained
in $\cA_{0-}$. We are now ready to prove the following theorem.

\begin{thm} \label{bfromu(b1)}
Let $\varphi$ be the injectiv linear map from $\mathcal{B}_{-1}$ to $\mathcal{U}(\mathcal{B}_1)_{-1}$ given by
\begin{align}
\varphi({F}^\cN) : E_\cM \mapsto {\rm ad }^\ast\,{\{E_\cM,F^\cN\}}.
\end{align}
Then $\cB$ is isomorphic to the subalgebra $\mathcal{V}\,'(\mathcal{B}_1,\varphi(\cB_{-1}))$ of $\mathcal{U}(\cB_1)$.
\end{thm}

\Pf
We can extend $\varphi$ to a homomorphism $\mathcal{A}_- \to V_-$, where $V=\mathcal{V}(\mathcal{B}_1,\varphi(\cB_{-1}))$.
Then it can be shown, in the same way as in \cite{Kantor-graded} (Lemma 5) that $\text{ker }\varphi$ is the maximal ideal of $\mathcal{A}$
contained in $\mathcal{A}_{0-}$, and thus
$\text{ker }\varphi = \mathcal{C}_-$.
It follows that
$\mathcal{A}_{0-}/\mathcal{C}_-$ is isomorphic to $V_-$. Furthermore, $\cA_+=V_+$, since both are generated freely by $\cA_1=\cB_1$, so 
$\mathcal{A}/\mathcal{C}_-$ is isomorphic to $V$.
It is clear that any graded ideal of $\mathcal{A}$ contained in $\mathcal{A}_{2+}$
is also an ideal of $\mathcal{A}/\mathcal{C}_-$ (via the natural embedding of $\mathcal{A}/\mathcal{C}_-$ into $\mathcal{A}$),
and also the other way around. By the definition of 
$V'=\mathcal{V}\,'(\mathcal{B}_1,\varphi(\cB_{-1}))$, it follows
that $V'$ is isomorphic to $(\mathcal{A}/\mathcal{C}_-)/\mathcal{C}_+$, which in turn is isomorphic to
$\mathcal{A}/(\mathcal{C}_-\oplus\mathcal{C}_+)=\mathcal{B}$.
\qed

\section{The tensor hierarchy algebra} \label{tha-sec}

In the preceding subsection we considered $\mathcal{V}\,'(\mathcal{B}_1,V_{-1})$ in the case where $V{}_{-1}$ is a subspace of
$\mathcal{U}(\mathcal{B}_1)_{-1}$ isomorphic to $\mathcal{B}_{-1}$, with $\varphi$ as the vector space isomorphism (and in fact also
isomorphic to $\mathcal{B}_1$, since there is an isomorphism between $\cB_{-1}$ and $\cB_1$).
In that case the Lie algebra $\{V_{-1},\cB_1\}$ is isomorphic to a direct sum of $\g$ and a one-dimensional Lie algebra
(corresponding to the element $h$).
Now instead let $V_{-1}$ be a subspace of $\mathcal{U}(\mathcal{B}_1)_{-1}$
such that 
\begin{itemize}
\item[{(i)}] the Lie algebra $\{V_{-1},\cB_1\}$ is isomorphic to $\g$,
\item[{(ii)}] satisfies the condition $\mathcal{V}\,'(\mathcal{B}_1,V_{-1})_2 \subseteq \cB_2$, and
\item[{(iii)}] contains at least one element $\Theta$ such that $\{\Th,\Th\}=0$ and the subspace $\{\Th,\cB_1\}$ is a non-abelian subalgebra of $\g$.
\end{itemize}
Let $\cT_{-1}$ be the direct sum of all such
$V_{-1} \subseteq \mathcal{U}(\mathcal{B}_1)_{-1}$, and set $\mathcal{V}\,'(\mathcal{B}_1,\cT_{-1})=\cT$.
Then we call $\cT$ a {\it tensor hierarchy algebra}.

We will show that when $\g=\mathfrak{e}_{11-D}$
the representations ${\bf t}_p$ of
$\g$ on $\cT_p$ given by the adjoint action,
\begin{align}
{\bf t}_p : \quad \g \to \End\cT_p,\quad {\bf t}_p(x)(y)=\dlb x,y \drb.
\end{align}
for all $p \geq 1$, are
precisely the representations ${\bf r}_p$ that arise in the tensor hierarchy of gauged maximal supergravity in $D$ dimensions.
We recall that these representations are given for $p \geq 2$ by the lower indices of the intertwiners
$Y^{\cM_1\cdots\cM_p}{}_{\cN_1\cdots\cN_{p+1}}$, which in turn are defined by the recursion formula
(\ref{y-relation}) and the initial condition (\ref{lilla1}),
taking into account the supersymmetry constraint on the embedding tensor.
The representation that we start with, for $p=1$, is always the same for the tensor hierarchy, the BKM superalgebra $\cB$,
and the tensor hierarchy algebra $\cT$, that is
${\bf r}_1={\bf s}_1={\bf t}_1$. What we will show is that $\cT$ gives exactly the tensor hierarchy representations,
and that these are contained in the representations coming from $\cB$,
that is ${\bf r}_p  = {\bf t}_p \subseteq {\bf s}_p$ for
all $p \geq 1$. (The use of the notation ${\bf r}_p$, ${\bf s}_p$, ${\bf t}_p$ here should note be confused with those in
\cite{Palmkvist:2011vz,Palmkvist:2012nc}.)

As the reader might have guessed, for $\g=\mathfrak{e}_{11-D}$ the element $\Theta$ above corresponds to the embedding tensor, and the subalgebra $\{\Th,\cB_1\}$
of $\g$ is the Lie algebra of the gauge group ${\rm G}_0 \subseteq {\rm G}$. Furthermore, we will see that
the conditions $\cT_2 \subseteq \cB_2$ and $\{\Th,\Th\}=0$ precisely encode the supersymmetry constraint and the closure constraint, respectively.
Thus the condition $\{\Th,\Th\}=0$ in the third assumption above already implies that $\{\Th,\cB_1\}$ is a subalgebra of $\g$, so we only add the condition that it be non-abelian. The condition $\cT_2 \subseteq \cB_2$ will be referred to as the supersymmetry constraint also in the general case, although 
the connection to supersymmetry is so far established only for $\g=\mathfrak{e}_{11-D}$. We will have a closer look at it in the next subsection.

\subsection{The supersymmetry constraint} \label{susyconstrsec}

First, let $V_{-1}$ be a subspace of $\mathcal{U}(\mathcal{B}_1)_{-1}$ that is only required to satisfy the first condition the preceding subsection, that
the Lie algebra $\{V_{-1},\cB_1\}$ be isomorphic to $\g$, without containing a one-dimensional subalgebra commuting with $\g$. As before, set
$V=\mathcal{V}(\mathcal{B}_1,V_{-1})$.
We can introduce a basis of $V_{-1}$, where the basis elements $\Phi^\cM{}_\al$ are defined by
\begin{align}
\Phi^\cM{}_\al(E_\cN) = \{ E_\cN ,\Phi^\cM{}_\al\} = \de_\cN{}^\cM t_\al.
\end{align}
As the index structure indicates, and as can be checked by computing $[ t^\alpha,\, \Phi^\cM{}_\be ]$ in $V$, the representation of $\g$ on
$V_{-1}$ is $\overline{\bf r}_1\times{\bf adj}$. Without computing this tensor product explicitly we can distinguish two irreducible
representations that it always contains. First,
$\overline{\bf r}_1$ itself, with the interwiner $(t^\al)_\cM{}^\cN$, and second, the irreducible representation whose highest weight is the sum of the highest weights of $\overline{\bf r}_1$ and ${\bf adj}$.
We will 
denote them by ${\bf r}_{\rm S}$ and ${\bf r}_{\rm L}$, respectively, and
the orthogonal complement
to ${\bf r}_{\rm S} + {\bf r}_{\rm L}$ in 
$\overline{\bf r}_1\times{\bf adj}$ (which need not be irreducible)
by ${\bf r}_{\rm M}$.
For $\g=\frake_{11-D}$, the representations ${\bf r}_{\rm S}$, ${\bf r}_{\rm M}$, ${\bf r}_{\rm L}$ are given by the table below.
\setlength{\arraycolsep}{14.75pt}
\\
{\renewcommand{\arraystretch}{1.5}
\begin{align*}
\begin{array}{|c|c|c|c|c|c|c|}
\hline
D&\g&{\bf r}_1&{\bf adj}&\overline{\bf r}_1={\bf r}_{\rm S}&{\bf r}_{\rm M}&{\bf r}_{\rm L}\\
\hline
7&\mathfrak{a}_4&\overline{\bf 10}&{\bf 24}&{\bf 10}&{\bf 15}+\overline{\bf 40}&{\bf 175}\\
6&\mathfrak{d}_5 &{\bf 16}_c & {\bf 45} &{\bf 16}_s&{\bf 144}_c&{\bf 560}_s\\
5&\mathfrak{e}_6  &\overline{\bf 27}  & {\bf 78} &{\bf 27}&\overline{\bf 351}&{\bf 1728}\\
4&\mathfrak{e}_7 &{\bf 56} & {\bf 133} &{\bf 56}&{\bf 912}&{\bf 6480}\\
3&\mathfrak{e}_8 &{\bf 248}  & {\bf 248} &{\bf 248}&{\bf 1}+{\bf 3875}+{\bf 30380}&{\bf 27000}\\
\hline
\end{array}
\end{align*}
}

For $\g=\frake_{11-D}$ and $D=3,4,5$, the highest weight of ${\bf r}_{\rm M}$ is the highest weight of $\overline{\bf r}_1$ plus the highest
root of the $\g=\frake_{10-D}$ subalgebra obtained by removing the
node in the Dynkin diagram of $\g$ to which the black one is attached in the Dynkin diagram of $\cB$. When $D=7$ this leads to a nonsimple Lie algebra
$\frake_3=\mathfrak{a}_1 \oplus \mathfrak{a}_2$, and ${\bf r}_{\rm M}$ is the direct sum of two irreducible representations, corresponding to the two simple
subalgebras $\mathfrak{a}_1$ and $\mathfrak{a}_2$. When $D=3$ we get ${\bf 3875}$ from the $\frake_7$ subalgebra of $\frake_8$, but as can be seen in the table there are two additional irreducible representations in ${\bf r}_{\rm M}$. Since in fact one of them is smaller than ${\bf r}_{\rm S}$ and the other is larger than
${\bf r}_{\rm L}$, the notation ${\bf r}_{\rm S}$, ${\bf r}_{\rm M}$, ${\bf r}_{\rm L}$ is not really appropriate in this case. Alternatively, one can
decompose the tensor product ${\bf 248} \times {\bf 248}$ into its symmetric and antisymmetric parts,
\begin{align}
({\bf 248} \times {\bf 248})_+ &= {\bf 1}+{\bf 3875}+{\bf 27000},\nn\\
({\bf 248} \times {\bf 248})_- &= {\bf 248}+{\bf 30380}.
\end{align}

We go back to the general case, and the direct sum $\overline{\bf r}_1\times{\bf adj} = {\bf r}_{\rm S} + {\bf r}_{\rm M} + {\bf r}_{\rm L}$.
The question is how much of this reducible representation
is included in ${\bf t}_{-1}$, or in other words, how much of $V_{-1}$ we can include in $\cT_{-1}$
without violating the condition $\cT_{2}\subseteq \cB_2$.
We recall that $V'=\mathcal{V}\,'(\mathcal{B}_1,V_{-1})$ is obtained from
$V=\mathcal{V}(\mathcal{B}_1,V_{-1})$ by factoring out the maximal graded ideal contained in $V_{2+}$. 
In view of Theorem \ref{bfromu(b1)}, 
the condition $\cT_2 \subseteq \cB_2$ can be put differently:
the part of $V_2$ that commutes with $\cT_{-1}$ must contain the part of $V_2$ that commutes with $\varphi({F}^\cM)$
for all $F^\cM$. Thus we need to study also the representation of $V_0=\g$ on $V_2$ given by the adjoint action, and the irreducible parts that it decomposes into.
Since $V_+$ is freely generated by $V_1=\cB_1$, this 
is the symmetric tensor product of two ${\bf r}_1={\bf s}_1$ representations.
It decomposes into a direct sum of ${\bf s}_2$ (corresponding to the part $\cB_2 \subset V_2$),
and its orthogonal complement, which we denote by ${\bf s}_\perp$. Thus
\begin{align}
({\bf r}_1 \times {\bf r}_1)_+ = {\bf s}_2 + {\bf s}_\perp.
\end{align}
Since $[\Phi^\cM{}_\al,\{ E_\cN,\,E_\cP \}]= 2\de_{(\cN}{}^\cM (t_\al)_{\cP)}{}^{\cQ}E_\cQ$, we need to study the expression
\begin{align} \label{studieuttryck}
\mathbb{P}^\cM{}_\al{}_\cR{}^\be \, (\mathbb{P}_\perp)_{\cN\cP}{}^{\cS\cT}\,\de_{\cS}{}^\cR (t_\be)_{\cT}{}^{\cQ},
\end{align}
where $\mathbb{P}$ is the projector corresponding to an irreducible representation ${\bf r}$ contained in the tensor product $\overline{\bf r}_1\times{\bf adj}$,
and
$\mathbb{P}_{\perp}$
is the projector corresponding to ${\bf s}_{\perp}$. 

In appendix \ref{superapp} we show that ${\bf r}_{\rm S}$ and ${\bf r}_{\rm L}$ cannot be contained in ${\bf t}_{-1}$, and if $D=3$ (so that 
${\bf r}_1=\overline{\bf r}_1={\bf adj}$), we show that ${\bf t}_{-1}$ must be contained in the symmetric part of the tensor product 
$\overline{\bf r}_1 \times {\bf adj}= {\bf r}_1 \times {\bf r}_1$. For $\g=\frake_{11-D}$, these two conditions turn out to
be equivalent to $\cT_2 \subseteq \cB_2$ and thus determine ${\bf t}_{-1}$ completely. It remains to see whether this
equivalence holds also in the general case.

Thus for $\g=\frake_{11-D}$ and $4 \leq D \leq 7$ we have ${\bf t}_{-1}={\bf r}_{\rm M}$, whereas for $\g=\frake_8$ the second condition removes the 
${\bf 30380}$ in ${\bf r}_{\rm M}$ and leaves us with ${\bf t}_{-1}={\bf 1}+{\bf 3875}$.
This result can be verified explicitly by inserting the expressions in appendix \ref{explapp} into (\ref{studieuttryck}) above.
Remarkably, it agrees with the constraint that supersymmetry puts on the embedding tensor in maximal gauged supergravity,
so that the embedding tensor always transforms in the representation ${\bf t}_{-1}$ for $\g=\frake_{11-D}$.

\subsection{The tensor hierarchy again} \label{tenshiersec}

Knowing ${\bf t}_{-1}$
we can compute the representations ${\bf t}_p$ for any $p \geq 2$.
The subspace $\cT_p$ is spanned by elements
\begin{align}
E_{\cM_1\cdots\cM_p} = \dlb E_{\cM_1}, \dlb E_{\cM_2}, \ldots,  \dlb E_{\cM_{p-1}}, E_{\cM_p} \drb \ldots \drb \drb.
\end{align}
To see which subrepresentations are present in ${\bf t}_p \subseteq ({\bf r}_1)^p$, we write
\begin{align} \label{theq}
\dlb \Phi^{\lceil\cM}{}_{\al\rfloor},\, E_{\cN_1\cdots\cN_p}\drb &= T^{\cM}{}_{\al\,\cN_1\cdots\cN_p}{}^{\cP_1\cdots\cP_{p-1}} E_{\cP_1\cdots\cP_{p-1}},
\end{align}
where $\Phi^{\lceil\cM}{}_{\al\rfloor}$ denotes $\Phi^{\cM}{}_{\al}$ projected on ${\bf t}_{-1}$.
The lower indices 
$\cN_1,\cN_2,\ldots,\cN_p$ of the structure constants
on the right hand side of (\ref{theq}) determine the representation ${\bf t}_p$ in the following sense. If (and only if) the expression is zero when the indices are projected on some
subrepresentation of $({\bf r}_1)^p$, then it is not present in ${\bf t}_p$, since otherwise $E_{\cM_1\cdots\cM_p}$ projected on this representation would generate
a non-zero ideal of $\cT$ contained in $\cT_{2+}$.

Elaborating the left hand side of (\ref{theq}), we get, using the Jacobi identity,
\begin{align}
\dlb \Phi^{\lceil\cM}{}_{\al\rfloor},\, E_{\cN_1\cdots\cN_p}\drb &= \dlb \{ \Phi^{\lceil\cM}{}_{\al\rfloor},\, E_{\cN_1} \} , E_{\cN_2\cdots\cN_p}\drb
- \dlb E_{\cN_1},\, \dlb \Phi^{\lceil\cM}{}_{\al\rfloor},\,E_{\cN_2\cdots\cN_p}\drb\drb\nn\\
&=\de_{\cN_1}{}^{\lceil\cM} (t_{\al\rfloor})_{\cN_2\cdots\cN_p}{}^{\cP_1\cdots\cP_{p-1}} E_{\cP_1\cdots\cP_{p-1}}\nn\\
&\quad\,- T^{\cM}{}_{\al\,\cN_2\cdots\cN_p}{}^{\cP_2\cdots\cP_{p-1}} E_{\cN_1\cP_2\cdots\cP_{p-1}}\nn\\
&=\big( \de_{\cN_1}{}^{\lceil\cM} (t_{\al\rfloor})_{\cN_2\cdots\cN_p}{}^{\cP_1\cdots\cP_{p-1}}\nn\\
&\quad\,\ \,- \de_{\cN_1}{}^{\cP_1} T^{\cM}{}_{\al\,\cN_2\cdots\cN_p}{}^{\cP_2\cdots\cP_{p-1}} \big) E_{\cP_1\cP_2\cdots\cP_{p-1}}.
\end{align}
For the structure constants that determine the representations ${\bf t}_p$ we thus end up with the recursion formula 
\begin{align} \label{y-relation0}
T^{\cM}{}_{\al\,\cN_1\cdots\cN_p}{}^{\cP_1\cdots\cP_{p-1}} =
\de_{\cN_1}{}^{\lceil\cM} (t_{\al\rfloor})_{\cN_2\cdots\cN_p}{}^{\la\cP_1\cdots\cP_{p-1}\ra}
- \de_{ \cN_1}{}^{\la\cP_1|} T^{\cM}{}_{\al\,\cN_2\cdots\cN_p }{}^{|\cP_2\cdots\cP_{p-1}\ra} 
\end{align}
(where the angle brackets denote projection on ${\bf t}_{p-1}$),
and since
\begin{align}
[\Phi^\cM{}_\al,\{E_\cN,E_\cP\}]=2 \de_{(\cN}{}^\cM(t_\al)_{\cP)}{}^\cQ E_\cQ
\end{align}
we have
the initial condition
\begin{align} \label{lilla10}
T^\cM{}_{\al\,\cN\cP}{}^\cQ = 2\,\de_{(\cN}{}^{\lceil \cM} (t_{\al\rfloor})_{\cP)}{}^\cQ.
\end{align}
For $\g=\frake_{11-D}$ we can contract $T^{\cM}{}_{\al\,\cN_1\cdots\cN_p}{}^{\cP_1\cdots\cP_{p-1}}$ with the embedding tensor
$\Th_\cM{}^\al$
in the first two indices. Since the embedding tensor transforms in ${\bf t}_{-1}$ we do not loose any representation by this contraction. If we now set 
\begin{align}
Y^{\cN_1\cdots\cN_p}{}_{\cM_1\cdots\cM_{p+1}} = \Theta_\cM{}^\al T^{\cM}{}_{\al\,\cM_1\cdots\cM_{p+1}}{}^{\cN_1\cdots\cN_{p}}
\end{align}
we obtain exactly (\ref{y-relation}) and (\ref{lilla1})
from (\ref{y-relation0}) and (\ref{lilla10}),
proving that ${\bf t}_p={\bf r}_p$ for all $p \geq 1$. We will henceforth only use the notation
${\bf r}_p$ for these representations, and extend it by setting ${\bf r}_p = {\bf t}_p$ also for $p\leq 0$.

Thus the embedding tensor can be considered as an element $\Theta=\Theta_\cM{}^\al \Phi^\cM{}_\al \in \cT_{-1}$, and the
intertwiners $Y^{\cN_1 \cdots \cN_{p}}{}_{\cM_1 \cdots \cM_{p+1}}$
can be considered as the components of the odd superderivation $(\text{ad\,}\Theta)$, which acts as
\begin{align}
\dlb \Theta, E_{\cM_1 \cdots \cM_{p+1}} \drb = Y^{\cN_1 \cdots \cN_{p}}{}_{\cM_1 \cdots \cM_{p+1}} E_{\cN_1 \cdots \cN_{p}}.
\end{align}
In section \ref{closconstrsec} we will see that the closure constraint forces this element to square to zero, $\{\Theta,\Theta\}=0$.

We end this subsection by showing that ${\bf r}_p \subseteq {\bf s}_p$ for all $p \geq 1$.
Both $\cB_+$ and $\cT_+$ are subalgebras of $V_+$ (the free Lie superalgebra generated by $\cB_1$) obtained by factoring out ideals,
\begin{align}
\cB_+ &= V_+/\cC_+, & \cT_+ &= V_+/\cD_+,
\end{align}
where $\cC_+$ is generated by $\{e_0,e_0\}$ (see section \ref{constrfruglsa}), and
$\cD_+$ is the maximal graded ideal of $V=\cV(\cB,V_{-1})$ contained in $V_{2+}$. The condition
$\cT_2 \subseteq \cB_2$ means $(\cC_+)_2 \subseteq (\cD_+)_2$. Since $\cC_+$ is generated by $(\cC_+)_2$, we have
$(\cC_+)_p \subseteq (\cD_+)_p$ and thus ${\bf r}_p \subseteq {\bf s}_p$ for all $p \geq 2$.

As long as ${\bf s}_p$ is irreducible, we have ${\bf r}_p = {\bf s}_p$,
but otherwise we may have ${\bf r}_p \neq {\bf s}_p$.
In the case of maximal supergravity in $D=3$ dimensions ($\g=\mathfrak{e}_8$)
this happens already for $p=2$, with ${\bf r}_2={\bf 3875}$ and ${\bf s}_2={\bf3875}+{\bf 1}$. This in turns leads to a
${\bf 248}$ in ${\bf s}_3$ which is missing in ${\bf r}_3$.
But this is nothing special for $D=3$, also for $D=5$ and $p=6$ the tensor hierarchy misses a singlet,
and we suspect that there are more examples for other $D$ and sufficiently large $p$.

\subsection{The end of the hierarchy} \label{end-sec}

We will now see what happens when we approach, and exceed, the spacetime limit $p=D$, which is the end of the tensor hierarchy in the sense that
$p$-form fields with $p > D$ are zero. (The title of this subsection is inspired by \cite{deWit:2008gc}.)

Because of the connection between $\cB$ and the affine Kac-Moody algebra $\g_{\rm KM}$ described in the section \ref{cs-constr}, there is an
`affine structure' also in $\cB$ in the sense that ${\bf s}_{D-2}={\bf adj}$, and ${\bf s}_{D-2-p}$ is conjugate to ${\bf s}_p$ for $1 \leq p \leq D-3$.
This was explained in more detail in \cite{Kleinschmidt:2013em} (explicitly for $\g=\frake_{11-D}$) using the results of \cite{Palmkvist:2012nc}.
In particular, ${\bf s}_{D-3}=\overline{\bf s}_1$, and there is a basis $E^\cM$ of $\cB_{D-3}$ with a normalization such that
$\dlb E_\cM,E^\cN \drb =(t_\al)_\cM{}^\cN E^\al$, where $E^\al$ is a basis of $\cB_{D-2}$ (and the same holds in $\cT$).
Now we have
\begin{align}
\dlb \{ E_\cN,E_\cP \},E^\cQ\drb = 2 (t_\be)_{(\cP}{}^\cQ \dlb E_{\cN)},E^\be \drb.  
\end{align}
As in section \ref{susyconstrsec}, let ${\bf r}$ be an irreducible representation contained in the tensor product $\overline{\bf r}_1\times{\bf adj}$, with a corresponding projector $\mathbb{P}$ (acting from the left). Then the conjugate $\overline{\bf r}$ is contained in ${\bf r}_1\times{\bf adj}$, and $\mathbb{P}$ acting from the right corresponds to $\overline{\bf r}$. In order to see whether $\overline{\bf r}$ is allowed in ${\bf t}_{D-1}$ or not, we thus need to
study the expression
\begin{align} \label{studieuttryck2}
(\mathbb{P}_\perp)_{\cN\cP}{}^{\cS\cT} (t_\be)_{\cT}{}^{\cQ} \de_{\cS}{}^\cR \,\mathbb{P}^\cM{}_\al{}_\cR{}^\be
\end{align}
If this is not zero, then the part of $\dlb E_{\cM},E^\al \drb$ corresponding to $\overline{\bf r}_{}$ must be zero, since
$(\mathbb{P}_\perp)_{\cN\cP}{}^{\cS\cT}\{E_\cS,E_\cT\}=0$.
As we saw in section \ref{susyconstrsec}, the representations ${\bf r}$ for which (\ref{studieuttryck2}) is zero are (by definition)
exactly the irreducible representations that are included in ${\bf r}_{-1}={\bf t}_{-1}$, so we have
${\bf s}_{D-1} \subseteq \overline{\bf r}_{-1}$. (In fact ${\bf s}_{D-1} = \overline{\bf r}_{-1}$ for $\g=\frake_{11-D}$, and probably also in the general case.) 
Since in turn ${\bf r}_{D-1} \subseteq {\bf s}_{D-1}$ we conclude that 
${\bf r}_{D-1}$, the representation in which the $(D-1)$-form field transforms, is contained in $\overline{\bf r}_{-1}$,
the conjugate of the representation in which the embedding tensor transforms.

\subsection{Beyond the end} \label{beyendsec}

For the subspaces $\cT_{D-2+p}$ with $p \geq 1$ we can introduce a basis
\begin{align}
E_{\cM_1\cdots\cM_p}{}^\al = \dlb E_{\cM_1}, \dlb E_{\cM_2}, \ldots,  \dlb E_{\cM_{p-1}}, \dlb E_{\cM_p}, E^\al \drb \drb \ldots \drb \drb.
\end{align}
Just replacing the indices in section \ref{tenshiersec} we can then write
\begin{align}
\dlb \Phi^{\lceil\cM}{}_{\al\rfloor},\,E_{\cP_1 \cdots \cP_p}{}^\ga \drb 
&= T^\cM{}_{\al\,\cP_1 \cdots \cP_p}{}^{\ga\,\cN_1\cdots\cN_{p-1}}{}_\be \, E_{\cN_1\cdots\cN_{p-1}}{}^\be,
\end{align}
and elaborate on the left hand side,
\begin{align}
\dlb \Phi^{\lceil\cM}{}_{\al\rfloor},\,E_{\cP_1 \cdots \cP_p}{}^\ga \drb 
&= [\{\Phi^{\lceil\cM}{}_{\al\rfloor},\,E_{\cP_1}\},\,E_{\cP_2 \cdots \cP_p}{}^\ga] \nn\\
&\quad\,- \dlb E_{\cP_1},\, \dlb \Phi^{\lceil\cM}{}_{\al\rfloor},\,E_{\cP_2\cdots\cP_{p-1}}{}^\ga \drb \drb \nn\\
&=\de_{\cP_1}{}^{\lceil\cM} (t_{\al\rfloor})_{\cP_2 \cdots \cP_p}{}^{\ga\,\cN_1\cdots\cN_{p-1}}{}_\be \, E_{\cN_1\cdots\cN_{p-1}}{}^\be\nn\\
&\quad\,- \de_{\cP_1}{}^{\cN_1} T^\cM{}_{\al\,\cP_2 \cdots \cP_p}{}^{\ga\,\cN_2\cdots\cN_{p-1}}{}_\be \,E_{\cN_1\cN_2\cdots\cN_{p-1}}{}^\be,
\end{align}
so that we get the recursion formula
\begin{align} \label{M-recursion}
T^\cM{}_{\al\,\cP_1 \cdots \cP_p}{}^{\ga\,\cN_1\cdots\cN_{p-1}}{}_\be&=
\de_{\cP_1}{}^{\lceil\cM} (t_{\al\rfloor})_{\cP_2 \cdots \cP_p}{}^{\ga\,\cN_1\cdots\cN_{p-1}}{}_\be \nn\\
&\quad\,- \de_{\cP_1}{}^{\lceil \cN_1|} T^{\cM}{}_{\al\,\cP_2 \cdots \cP_p}{}^{\ga\,|\cN_2\cdots\cN_{p-1}}{}_{\be\rfloor}
\end{align}
for $p \geq 3$, where the `diagonal' brackets around the indices 
$\cN_1,\cN_2\ldots,\cN_{p-1},{\be}$
denote projection on $\overline{\bf r}_{D-2+(p-1)}$. 
For $p=2$ and $p=1$ we similarly have
\begin{align} \label{M-recursion2}
T^\cM{}_{\al\,\cP \cQ}{}^{\ga\,\cN}{}_\be&=\de_{\cP}{}^{\lceil\cM} (t_{\al\rfloor})_{\cQ}{}^{\ga\,\cN}{}_\be 
- \de_{\cP}{}^{\lceil \cN|} T^{\cM}{}_{\al\,\cQ}{}^{\ga}{}_{|\be\rfloor},\nn\\
T^\cM{}_{\al\,\cP}{}^{\ga}{}_\be&=\de_\cP{}^{\lceil\cM} f_{\al\rfloor}{}^\ga{}_\be - T^\cM{}_\al{}^\ga{}_\cN (t_\be)_\cP{}^\cN,
\end{align}
where $T^{\cM}{}_{\al\,\cQ}{}^{\ga}{}_{\be}$ and $T^\cM{}_\al{}^{\be}{}_\cN$ are defined by
\begin{align}
\dlb \Phi^{\lceil\cM}{}_{\al\rfloor}, E{}_\cQ{}^\ga \drb &= T^{\cM}{}_{\al\,\cQ}{}^{\ga}{}_{\be} E^\be, &
\dlb \Phi^{\lceil\cM}{}_{\al\rfloor}, E^\be \drb &= T^\cM{}_\al{}^{\be}{}_\cN E^\cN.
\end{align}
By induction it is clear that all the structure constants $T^{\cM}{}_{\al\,\cN_1\cdots\cN_p}{}^{\cP_1\cdots\cP_{p-1}}$ are invariant tensors of
$\g$ (also when we change basis of $\cT_p$ for $p \geq D-3$), and then $T^\cM{}_\al{}^{\be}{}_\cN$ must be a linear combination of the projectors of the irreducible representations contained in
${\bf r}_{D-1}$. In section \ref{closconstrsec} we will see that all the coefficients in this linear combination are in fact equal to one,
$T^\cM{}_\al{}^{\be}{}_\cN = \de_\cN{}^{\lceil \cM}\de_{\al \rfloor}{}^\be$.
With this equation as an initial condition, the recursion formulas (\ref{M-recursion})--(\ref{M-recursion2}) determine all the structure constants
$T^\cM{}_{\al\,\cP_1 \cdots \cP_p}{}^{\ga\,\cN_1\cdots\cN_{p-1}}{}_\be$, which in turn determine the representations 
${\bf r}_{(D-2)+p}$, not only for $p=1$ and $p=2$, but also for arbitrarily large $p$, beyond the end of the tensor hierarchy.

\subsection{Beyond the beginning}

As in section \ref{susyconstrsec},
we let $V_{-1}$ be a general subspace of $\mathcal{U}(\mathcal{B}_1)_{-1}$ such that
the Lie algebra $\{V_{-1},\cB_1\}$ is isomorphic to $\g$.
We can extend the basis $\Phi^\cM{}_\al$ of $V_{-1}$ to a basis of $V_-$, 
where $\Phi^{\cM_1 \cdots \cM_{p}}{}_{\al}$ is a basis of the subspace $V_{-p}$ for $p \geq 1$,
defined recursively by
\begin{align}
\Phi^{\cM_1 \cdots \cM_{p}}{}_{\al} (E_\cN) = \de_\cN{}^{\cM_1} \Phi^{\cM_2 \cdots \cM_{p}}{}_{\al}.
\end{align}
Let us furthermore denote projection of $\Phi^{\cM_1 \cdots \cM_{p}}{}_{\al}$ on ${\bf r}_{-p}$ by $\Phi^{\lceil\cM_1 \cdots \cM_{p}}{}_{\al\rfloor}$, so that
this is a basis of $\cT_{-p} \subset V_{-p}$, and define the structure constants $S^\cM{}_{\al}{}^{\cN_1 \cdots \cN_{p-1}}{}_{\be}{}_{\cP_1\cdots\cP_{p}}{}^\ga$ by
\begin{align}
\dlb \Phi^{\lceil\cM}{}_{\al\rfloor},\,\Phi^{\lceil\cN_1 \cdots \cN_{p-1}}{}_{\be\rfloor}   \drb
&= S^\cM{}_{\al}{}^{\cN_1 \cdots \cN_{p-1}}{}_{\be}{}_{\cP_1\cdots\cP_{p}}{}^\ga \, \Phi^{\cP_1\cdots\cP_{p}}{}_\ga.
\end{align}
Thus the structure of the contracted indices in this equation determines
the representation ${\bf r}_{-p}$.
We can obtain a recursion formula for the structure constants by first computing
\begin{align}
\dlb \Phi^{\lceil\cM}{}_{\al\rfloor},\,\Phi^{\cN_1 \cdots \cN_{p-1}}{}_{\be} \drb (E_{\cP_1})
&=\dlb E_{\cP_1},\,\dlb \Phi^{\lceil\cM}{}_{\al\rfloor},\,\Phi^{\cN_1 \cdots \cN_{p-1}}{}_\be \drb \drb \nn\\
&=[\{E_{\cP_1},\,\Phi^{\lceil\cM}{}_{\al\rfloor}\},\,\Phi^{\cN_1 \cdots \cN_{p-1}}{}_\be]\nn\\
&\quad\,- \dlb \Phi^{\lceil\cM}{}_{\al\rfloor} ,\, \dlb E_{\cP_1},\,\Phi^{\cN_1 \cdots \cN_{p-1}}{}_\be \drb \drb \nn\\
&=\de_{\cP_1}{}^{\lceil\cM}  \dlb t_{\al\rfloor},\,  \Phi^{\cN_1 \cdots \cN_{p-1}}{}_\be \drb \nn\\
&\quad\,- \de_{\cP_1}{}^{\cN_1}\dlb \Phi^{\lceil\cM}{}_{\al\rfloor},\,\Phi^{\cN_2 \cdots \cN_{p-1}}{}_\be \drb\nn\\
&=\de_{\cP_1}{}^{\lceil\cM}   (t_{\al\rfloor})^{\cN_1 \cdots \cN_{p-1}}{}_\be{}_{\cP_2 \cdots \cP_p}{}^\ga \Phi^{\cP_2 \cdots \cP_p}{}_\ga  \nn\\
&\quad\,-\de_{\cP_1}{}^{\cN_1}S^\cM{}_{\al}{}^{\cN_2 \cdots \cN_{p-1}}{}_{\be}{}_{\cP_2\cdots\cP_{p}}{}^\ga \, \Phi^{\cP_2\cdots\cP_{p}}{}_\ga,
\end{align}
for $p \geq 3$, which then gives
\begin{align} \label{N-recursion}
S^\cM{}_{\al}{}^{\cN_1 \cdots \cN_{p-1}}{}_{\be}{}_{\cP_1\cdots\cP_{p}}{}^\ga
&= \de_{\cP_1}{}^{\lceil\cM}   (t_{\al\rfloor})^{\cN_1 \cdots \cN_{p-1}}{}_\be{}_{\cP_2 \cdots \cP_p}{}^\ga\nn\\
&\quad\,-\de_{\cP_1}{}^{\lceil\cN_1|}S^\cM{}_{\al}{}^{|\cN_2 \cdots \cN_{p-1}}{}_{\be\rfloor}{}_{\cP_2\cdots\cP_{p}}{}^\ga\nn\\
&= -\,\de_{\cP_1}{}^{\lceil\cM}   (t_{\al\rfloor}){}_{\cP_2 \cdots \cP_p}{}^\ga{}^{\cN_1 \cdots \cN_{p-1}}{}_\be\nn\\
&\quad\,-\de_{\cP_1}{}^{\lceil\cN_1|}S^\cM{}_{\al}{}^{|\cN_2 \cdots \cN_{p-1}}{}_{\be\rfloor}{}_{\cP_2\cdots\cP_{p}}{}^\ga.
\end{align}
For $p=2$ we have
\begin{align}
\{\Phi^{\cM}{}_{\al},\Phi^{\cN}{}_{\beta}\}(E_\cP)&=
[ E_\cP, \{\Phi^{\cM}{}_{\al},\Phi^{\cN}{}_{\beta}\} ]\nn\\
&=[\{ E_\cP, \Phi^{\cM}{}_{\al}\} ,\Phi^{\cN}{}_{\be} ] +
[\{ E_\cP, \Phi^{\cN}{}_{\be}\} ,\Phi^{\cM}{}_{\al} ]\nn\\
&=\big(-\de_\cP{}^{\cM} (t_{\al})_\cQ{}^{\cN} \de_{\be}{}^\ga
+ \de_\cP{}^{\cM} f_{\al\be}{}^\ga \de_\cQ{}^{\cN}\nn\\
&\quad\ \,\,\,-\de_\cP{}^{\cN} (t_{\be})_\cQ{}^{\cM} \de_{\al}{}^\ga
+ \de_\cP{}^{\cN} f_{\be\al}{}^\ga \de_\cQ{}^{\cM} \big)\Phi^\cQ{}_\ga,
\end{align}
which gives
\begin{align} \label{closureinitial}
S^{\cM}{}_{\al}{}^{\cN}{}_{\be}{}_{\cP\cQ}{}^\ga &=
-\,\de_\cP{}^{\lceil\cM} (t_{\al\rfloor})_\cQ{}^{\lceil\cN} \de_{\be\rfloor}{}^\ga
+ \de_\cP{}^{\lceil\cM} f_{\al\rfloor\lfloor\be}{}^\ga \de_\cQ{}^{\cN\rceil}\nn\\
&\quad\,-\de_\cP{}^{\lceil\cN} (t_{\be\rfloor})_\cQ{}^{\lceil\cM} \de_{\al\rfloor}{}^\ga
+ \de_\cP{}^{\lceil\cN} f_{\be\rfloor\lfloor\al}{}^\ga \de_\cQ{}^{\cM\rceil}.
\end{align}
We can extend the notation by introducing $S^\cM{}_{\al\,\be\,\cP}{}^\ga$ and $S^\cM{}_{\al\,\cN}{}^\be$ defined by
\begin{align}
[\Phi^{\lceil\cM}{}_{\al\rfloor}, t_\be] &= S^\cM{}_{\al\,\be\,\cP}{}^\ga \Phi^\cP{}_\ga, &
\{\Phi^{\lceil\cM}{}_{\al\rfloor}, E_\cN \} &= -S^\cM{}_{\al\,\cN}{}^\be t_\be,
\end{align}
where we know that $S^\cM{}_{\al\,\cN}{}^\be = -\,\de_\cN{}^{\lceil\cM}\de_{\al\rfloor}{}^\be$ and
\begin{align} \label{J2formel}
S^\cM{}_{\al\,\be\,\cP}{}^\ga &= (t_\be)_\cP{}^{\lceil\cM} \de_{\al\rfloor}{}^\ga  + \de_\cP{}^{\lceil\cM} f_{\al\rfloor\be}{}^\ga.
\end{align}
Then (\ref{closureinitial}) and (\ref{J2formel}) can be written
\begin{align}
S^\cM{}_{\al}{}^{\cN}{}_{\be}{}_{\cP\cQ}{}^\ga &= -\,\de_{\cP}{}^{\lceil\cM}   (t_{\al\rfloor}){}_{\cQ}{}^\ga{}^{\cN}{}_\be
-\de_{\cP}{}^{\lceil\cN|}S^\cM{}_{\al}{}_{|\be\rfloor}{}_{\cQ}{}^\ga,\nn\\
S^\cM{}_{\al\,\be\,\cP}{}^\ga &=-\,\de_\cP{}^{\lceil\cM} f_{\al\rfloor}{}^\ga{}_\be - S^\cM{}_\al{}_\cN{}^\ga(t_\be)_\cP{}^\cN,
\end{align}
extending the recursion formula (\ref{N-recursion}) for the structure constants
$S^\cM{}_{\al}{}^{\cN_1 \cdots \cN_{p-1}}{}_{\be}{}_{\cP_1\cdots\cP_{p}}{}^\ga$ to $p=2$ and $p=1$, with the initial condition
$S^\cM{}_{\al\,\cN}{}^\be = -\de_\cN{}^{\lceil\cM}\de_{\al\rfloor}{}^\be$.

Comparing the results in this subsection with those in the preceding one, we find that
$S^\cM{}_{\al\,\cN}{}^\be = -T^\cM{}_\al{}^{\be}{}_\cN$, and that the structure constants
$S^\cM{}_{\al}{}^{\cN_1 \cdots \cN_{p-1}}{}_{\be}{}_{\cP_1\cdots\cP_{p}}{}^\ga$
satisfy the same recursion formula as $-T^\cM{}_{\al\,\cP_1 \cdots \cP_p}{}^{\ga\,\cN_1\cdots\cN_{p-1}}{}_\be$, for $p \geq 1$.
Thus we have
\begin{align}
S^\cM{}_{\al}{}^{\cN_1 \cdots \cN_{p-1}}{}_{\be}{}_{\cP_1\cdots\cP_{p}}{}^\ga = - T^\cM{}_{\al\,\cP_1 \cdots \cP_p}{}^{\ga\,\cN_1\cdots\cN_{p-1}}{}_\be
\end{align}
for all $p \geq 1$. 

Since the indices $\cP_1,\ldots,\cP_p,\ga$ of the structure constants $J^\cM{}_{\al}{}^{\cN_1 \cdots \cN_{p-1}}{}_{\be}{}_{\cP_1\cdots\cP_{p}}{}^\ga$
and $T^\cM{}_{\al\,\cP_1 \cdots \cP_p}{}^{\ga\,\cN_1\cdots\cN_{p-1}}{}_\be$ for $p\geq 2$ determine
the representations $\overline{\bf r}_{-p}$ and ${\bf r}_{(D-2)+p}$, respectively,  
we conclude that ${\bf r}_{-p}$ is the conjugate of ${\bf r}_{(D-2)+p}$ for all $p \geq 2$. Combined with the `affine structure' inherited from
the BKM superalgebra $\cB$, we thus have $\overline{\bf r}_p={\bf r}_{(D-2)-p}$ for all $p$ in the tensor hierarchy algebra $\cT$,
{\it except for $p=-1$}. In that case we only have $\overline{\bf r}_{-1} \subseteq {\bf r}_{D-1}$, and if ${\bf r}_{-1}$ is reducible
it may happen that the equality does not hold. The reason is that ${\bf r}_{D-1}$ is determined by
$T^\cM{}_{\al\,\cP}{}^{\ga}{}_\be$, whereas ${\bf r}_{-1}$ is not determined by $S^\cM{}_{\al\,\be\,\cP}{}^\ga$, but given by the supersymmetry constraint in the definition of the tensor hierarchy algebra $\cT$. However, one can easily see that the only possible part of ${\bf r}_{-1}$ that will not show up in
$S^\cM{}_{\al\,\be\,\cP}{}^\ga$ is a singlet. This is precisely what happens for $D=3$, $\g=\frake_8$, 
where ${\bf r}_{-1}={\bf 3875}$ and ${\bf r}_2={\bf 3875}+{\bf 1}$. Except for this possible singlet, we always have
$\overline{\bf r}_{-p}={\bf r}_{(D-2)-p}$ for all $p$.

\subsection{The closure constraint} \label{closconstrsec}

In the definition of the tensor hierarchy algebra $\cT$ we assumed that there is at least one element $\Theta$ in $\cT_{-1}$ such that
$\{\Th,\Th\}=0$, and $\{\Th,\cB_1\}$ is a non-abelian subalgebra of $\g$. Later we saw that we could indeed identify the components of
a general such element $\Th$ with the embedding tensor for $\g=\frake_{11-D}$. We will now show that these assumptions imply that
$T^\cM{}_\al{}^\be{}_\cN=\de_\cN{}^{\lceil \cM}\de_{\al \rfloor}{}^\be$, as we claimed in section \ref{beyendsec}. It suffices to show this in the case when
${\bf r}_{-1}$ is irreducible, and then apply the result to each irreducible part of a general ${\bf r}_{-1}$ separately. Thus, assume
$T^\cM{}_\al{}^\be{}_\cN=\chi\de_\cN{}^{\lceil \cM}\de_{\al \rfloor}{}^\be$ for some nonzero constant $\chi$.
Now we have on the one hand side
\begin{align}
\dlb \Phi^\cM{}_\al, \dlb \Phi^\cN{}_\be, E_\cP{}^\ga \drb \drb &=
T^\cN{}_\be{}_\cP{}^\ga{}_\de \dlb \Phi^\cM{}_\al, E^\de \drb 
=T^\cN{}_\be{}_\cP{}^\ga{}_\de T^\cM{}_\al{}^\de{}_\cQ E^\cQ\nn\\
&=\de_\cP{}^\cN f_\be{}^\ga{}_\de T^\cM{}_\al{}^\de{}_\cQ E^\cQ - T^\cN{}_\be{}^\ga{}_\cR (t_\de)_\cP{}^\cR T^\cM{}_\al{}^\de{}_\cQ E^\cQ\nn\\
&=\chi\,\de_\cP{}^\cN f_\be{}^\ga{}_\al E^\cM - \chi^2 (t_\al)_\cP{}^\cN \de_\be{}^\ga E^\cM,
\end{align}
which gives
\begin{align} \label{ekvmedx}
0&=[\{\Th,\Th\},E_\cP{}^\ga]=2\dlb \Th, \dlb \Th, E_\cP{}^\ga \drb \drb\nn\\ &= 2\,\Th_\cM{}^\al \Th_\cN{}^\be
\big(\chi\,\de_\cP{}^\cN f_\be{}^\ga{}_\al  - \chi^2 (t_\al)_\cP{}^\cN \de_\be{}^\ga \big) E^\cM\nn\\
&=2\big(\chi\,\Th_\cM{}^\al \Th_\cP{}^\be f_\be{}^\ga{}_\al  - \chi^2\, \Th_\cM{}^\al \Th_\cN{}^\ga(t_\al)_\cP{}^\cN\big)E^\cM.
\end{align}
On the other hand, contracting (\ref{closureinitial}) with $\Th_\cM{}^\al\Th_\cN{}^\be$ we obtain
\begin{align} \label{ekvutanx}
0&=[\{\Th,\Th\},E_\cP]\nn\\&=\Th_\cM{}^\al \Th_\cN{}^\be[\{\Phi^\cM{}_\al,\Phi^\cN{}_\be\},E_\cP]
\nn\\
&= \Th_\cM{}^\al \Th_\cN{}^\be S^\cM{}_\al{}^\cN{}_\be{}_{\cP\cQ}{}^\ga \Phi^\cQ{}_\ga\nn\\
&= 2\,\Th_\cM{}^\al \Th_\cN{}^\be 
\big(\de_\cP{}^\cN f_\be{}^\ga{}_\al  - (t_\al)_\cP{}^\cN \de_\be{}^\ga \big) \Phi^\cQ{}_\ga\nn\\
&=2\big(\Th_\cM{}^\al \Th_\cP{}^\be f_\be{}^\ga{}_\al - \Th_\cM{}^\al \Th_\cN{}^\ga(t_\al)_\cP{}^\cN\big)\Phi^\cQ{}_\ga.
\end{align}
Since we assume $\{\Th,\cT_1\}$ to be a non-abelian subalgebra of $\g$, we get
\begin{align}
\Th_\cM{}^\al \Th_\cN{}^\be f_{\al\be}{}^\ga t_\ga = [X_\cM,X_\cN] \neq 0,
\end{align}
and thus both of the two terms in (\ref{ekvmedx}) or (\ref{ekvutanx}) must be nonzero. It follows that we must have $\chi=1$, which was to be proven.

In fact, this is the only time we use the assumption that $\{\Th,\cT_1\}$ is a non-abelian subalgebra of $\g$. Otherwise it is not needed, but of course,
it is physically motivated. From a mathematical point of view, it would be desirable with a more general proof that $\chi=1$ (if it really is
true in the general case).
In many cases, one can conclude that $\chi=1$ just from a comparison between (\ref{ekvmedx}) and (\ref{ekvutanx}), and a study of the possible invariant tensors.

We recognize the expression within the parentheses in the last line of (\ref{ekvutanx}) from (\ref{closconstr}) as $\de_\cN(\Th_\cQ{}^\ga)$,
the variation of the embedding tensor with respect to $X_\cP$,
which is set to zero by the closure constraint. This is in accordance with
\begin{align}
\{\Th,\Th\} (E_\cP) &= [ E_\cP, \{\Th,\Th\} ] = 2 [\{E_\cP,\Th\},\Th] = 2 [X_\cP,\Th].
\end{align}
This expression transforms in the representation ${\bf r}_{-2}$, which, according to the results in the last subsection, is the conjugate of ${\bf r}_D$.

\section{Conclusion} \label{con-sec}

In this paper we have investigated two different ways of extending 
an arbitrary simple finite-dimensional Lie algebra $\g$ (given a certain extension of $\g$ to an affine Kac-Moody algebra $\g_{\rm KM}$)
to a Lie superalgebra with a $\mathbb{Z}$-grading and a corresponding level decomposition.
One way leads to the BKM algebra $\cB$, the other to the tensor hierarchy algebra $\cT$.

We have shown that $\cB$ can be constructed in two ways:
as the Lie superalgebra generated by the elements $e_I$ and $f_I$ modulo the Chevalley-Serre relations, and as the subalgebra
$\cV\,'(\cB_1,V_{-1})$ of the universal graded Lie superalgebra $\mathcal{U}(\cB_{1})$,
in the special case when $V_{-1}=\varphi(\cB_{-1})$.
By changing $V_{-1}$ appropriately in the second construction, we instead obtained the
tensor hierarchy algebra $\cT$.
A natural question is whether also the Chevalley-Serre construction can be modified in order to give the tensor hierarchy algebra $\cT$
instead of the BKM algebra $\cB$. As a first step towards such a construction of $\cT$ we 
can associate a generator $\phi_I$ to each node in the Dynkin diagram, in addition to the generators $e_I$ and $f_I$ of
$\cB$, so that $\cT$ is instead generated by $e_I$ and $\phi_I$. The relations that one would have to impose on these generators should then
differ from the Chevalley relations (\ref{chev-rel0}) in that $\dlb e_I,\phi_J \drb $ could be non-zero even if $I \neq J$.
A study of the commutation relations in $\cB$ among $e_I,f_I$ and the images 
of $e_i,f_i \in \g = \cT_0$ under an appropriate vector space isomorphism
\begin{align}
\cT_{-1} \oplus \cT_{0} \oplus \cT_{1} \to \cB_{-D+1} \oplus \cB_{-D+2} \oplus \cB_{-D+3}
\end{align} 
leads to the relations $[\phi_0,e_i] = 0$ if {and only if} $A_{0i}\neq0$, whereas $[e_0,\phi_i]=0$ for all $i$,
and of course, for the $\g$ subalgebra, $[e_i,\phi_j] =\delta_{ij}h_j$. 
Since there is no singlet in ${\bf r}_0={\bf adj}$ (unlike
${\bf s}_0={\bf adj}+{\bf 1}$), there cannot be an element $h_0$ in $\cT$, so we must have $\{e_0,\phi_0\}\in \g$.
We will not further develop the idea of a possible Chevalley-Serre-like construction of $\cT$ here, but leave to future research to work out exactly the relations that $e_I$ and $\phi_I$ have to satisfy (see also appendix \ref{superapp}).

We have shown that, in contrast to the relation $\overline{\bf s}_p = {\bf s}_{-p}$ for $\cB$, 
the representations ${\bf r}_p$ in the level decomposition of $\cT$
satisfy $\overline{\bf r}_p = {\bf r}_{(D-2)-p}$, up to a possible additional singlet in ${\bf r}_{-1}$.
Thus, on the subspace of $\cT$ obtained by removing this singlet (if present),
it must be possible to introduce a (super)involution $\tau$ and a non-degenerate bilinear form $\kappa$ such that
$\tau(\cT_p)=\cT_{(D-2)-p}$ and $\kappa(\cT_p,\cT_q)=0$ if $p+q \neq D-2$.
Hopefully, this can be done in a canonical way, giving properties of $\tau$ and $\kappa$ similar to those
of the Chevalley involution and the invariant bilinear form in a Kac-Moody algebra (or of their counterparts in 
the BKM superalgebra $\cB$). Such $\tau$ and $\kappa$ would probably be useful tools in order to investigate whether $\cT$ is really a symmetry of the
corresponding gauged supergravity theory, or just a device for keeping track of the representations that appear.
A related question is whether it is possible, for each $\g$, to unify the different tensor hierarchy algebras
$\cT$, corresponding to different $D$, into one `universal' tensor hierarchy algebra. As shown in \cite{Kleinschmidt:2013em},
this is possible for the BKM superalgebras $\cB$,
where `oxidation' from $D$ to $D+1$ corresponds to removing a node in an appropriate (non-distinguished) Dynkin diagram of $\cB$.

Finally, we mention that not only gauged supergravity theories, but also $D=3$ superconformal theories 
with $\cN=6$ or $\cN=5$ supersymmetry can be
characterized by graded Lie superalgebras $\cG$ of the form $\cV\,'(U_1,V_{-1})$. The
vector spaces $U_1$, $V_{-1}$ are then given by a three-algebra
\cite{Bagger:2007jr,Bagger:2008se,deMedeiros:2009eq,Chen:2009cwa,
Bagger:2010zq,Cantarini:2010kg,Gustavsson:2010yr,Palmkvist:2009qq,Kim:2010kq,Palmkvist:2011aw}
rather than by an embedding tensor, but the construction of the
graded Lie superalgebra from the vector spaces is the same.
An important difference is that $\cG$ unlike $\cT$ is finite-dimensional, but there is also a similarity in that
supersymmetry restricts the size of the subspaces on the positive side of the grading. For the $D=3$ superconformal theories 
the constraints of $\cN=6$ or $\cN=5$ supersymmetry are $\cG_2=0$ and $\cG_3=0$, respectively, corresponding to the supersymmetry
constraint $\cT_2 \subseteq \cB_2$ in gauged supergravity \cite{Palmkvist:2009qq,Kim:2010kq,Palmkvist:2011aw}.
Following \cite{Bergshoeff:2008cz} it was shown in \cite{Bergshoeff:2008bh} that three-dimensional superconformal theories with supersymmetry
$\cN=8$ or less can be obtained from the corresponding $D=3$ gauged supergravity theory
by taking a flat space limit. Within the framework of the present paper, it would be interesting to study this connection further.

\subsubsection*{Acknowledgments}

I would like to thank Martin Cederwall, Thibault Damour, Jesper Greitz, 
Jonas Hartwig, Marc Henneaux, Paul Howe, Bernard Julia, Victor Kac,
Axel Kleinschmidt, Carlo Meneghelli, Hermann Nicolai and Henning Samtleben
for discussions, answers to questions, and moral support.
I am also grateful to Nordita, Nordic Institute for Theoretical Physics, 
for hospitality during the finalization of this project.

\appendix

\section{More on the supersymmetry constraint} \label{superapp}

As promised in section \ref{susyconstrsec} we will in this appendix show that ${\bf r}_{-1} \subset {\bf r}_{\rm M}$, and, if $D=3$, 
that ${\bf r}_{-1} \subset ({\bf r}_1 \times {\bf r}_1)_+$.

Let $\mathfrak{h}^\ast$ be the root space of $\g$, and let $\Lambda$ be the map from $V$ to $\mathfrak{h}^\ast$
mapping a weight vector in any subspace $\cT_p \subset \cT$ (considered as a $\g$-module) on the corresponding weight.
Then $\Lambda(e_0)$ is the lowest weight of ${\bf r}_1$,
and $-\Lambda(e_0)$ is the highest weight of $\bar{\bf r}_1$. Furthermore, if $e_\theta$ is a highest root vector of $\g$, then
$-\Lambda(e_0)+\Lambda(e_\theta)$ is the highest weight of ${\bf r}_{\rm L} \subset \bar{\bf r}_1 \times {\bf adj}$
(and also the highest weight of $\bar{\bf r}_1 \times {\bf adj}$), which has multiplicity one. 
Let $\Phi$ be the map
\begin{align} 
\Phi:\cB_{-1} \times \g \to \cT_{-1}, \qquad \Phi(y,z)(x)=\{x,\Phi(y,z)\}=\langle x | y \rangle z,
\end{align}
where $x \in \cB_{1}=\cT_1$,
and $\la x | y \ra$ is the supersymmetric invariant bilinear form in $\cB$, for which we have $\langle E_\cM | F^\cN \rangle=\de_\cM{}^\cN$
\cite{Ray,Palmkvist:2012nc}, which means
that $\Phi(F^\cM,t_\al)=\Phi^\cM{}_\al$. One can easily show that $\Lambda(\Phi(y,z))=\Lambda(y)+\Lambda(z)$.
Then
\begin{align}
\Lambda(\Phi(f_0,e_\theta))=\Lambda(f_0)+\Lambda(e_\theta)=-\Lambda(e_0)+\Lambda(e_\theta),
\end{align}
and it follows that $\Phi(f_0,e_\theta)$ is a highest weight vector of ${\bf r}_{\rm L}$. Now we have
\begin{align}
[\Phi(f_0,e_\theta), \{ e_0,e_0\}] =2 [\{ \Phi(f_0,e_\theta), e_0 \},e_0] = 2 [e_\theta, e_0],
\end{align}
which is nonzero since
\begin{align}
\{ [e_\theta, e_0], f_0 \}=  [e_\theta, \{ e_0, f_0 \}]= -[h_0,e_\theta]\neq0.
\end{align}
Thus ${\bf r}_{\rm L}$ cannot be contained in
${\bf r}_{-1}$.

Considering ${\bf r}_{\rm S}$, set $\Phi^\cN=(t^\al)_\cM{}^\cN \Phi^\cM{}_\al$. The commutator with $\{E_\cP,E_\cQ\}$ is
\begin{align}
[\Phi^\cN,\{E_\cP,E_\cQ\}]=2 (t^\al)_{(\cP}{}^\cN (t_\al)_{\cQ)}{}^\cR E_\cR
\end{align}
so the question is whether the expression $(t^\al)_{(\cP}{}^\cN (t_\al)_{\cQ)}{}^\cR$
contains ${\bf s}_\perp$ in the lower indices or not. To answer this question, we compare $\Phi^\cN$ in $\cT_{-1}$ with $F^{\cN}$ in $\cB$,
for which we have
\begin{align}
[F^\cN,\{E_\cP,E_\cQ\}]=2\bigg( (t^\al)_{(\cP}{}^\cN (t_\al)_{\cQ)}{}^\cR + \frac1{D-2}\, \de_{(\cP}{}^\cN \de_{\cQ)}{}^\cR \bigg) E_\cR.
\end{align}
and we know that the expression within the parentheses contain only ${\bf s}_2$ in the lower indices. Since $\de_{(\cP}{}^\cN \de_{\cQ)}{}^\cR$
contains both ${\bf s}_2$ and ${\bf s}_\perp$, we conclude that $(t^\al)_{(\cP}{}^\cN (t_\al)_{\cQ)}{}^\cR$
must contain ${\bf s}_\perp$ as well, and thus ${\bf r}_{\rm S}$ cannot be contained in
${\bf r}_{-1}$.

For $D=3$ we write $\{E_\cP,\Phi^{\cM\cN}\}=\de_\cP{}^\cM t^\cN$ and $[t^\cN, E_\cQ]=f^\cN{}_{\cQ}{}^{\cR}E_\cR$, so we get
\begin{align}
[\Phi^{[\cM\cN]},\{E_\cP,E_\cQ\}]= \de_{(\cP}{}^{[\cM} f^{\cN]}{}_{\cQ)}{}^{\cR}E_\cR.
\end{align}
Contracting with $f^\cS{}_{\cM\cN}$ gives
\begin{align}
f^\cS{}_{\cM\cN}[\Phi^{\cM\cN},\{E_\cP,E_\cQ\}] &= f^\cS{}_{\cM\cN} \de_{(\cP}{}^\cM f^\cN{}_{\cQ)}{}^{\cR}E_\cR\nn\\
&= - f^\cS{}_{\cN(\cP} f^\cN{}_{\cQ)}{}^{\cR}E_\cR.
\end{align}
According to what we have just shown for a general $D$, the expression $f^\cS{}_{\cN(\cP} f^\cN{}_{\cQ)}{}^{\cR}$ contains
${\bf s}_\perp$ in the symmetric indices, and then this must hold for the uncontracted expression $\de_{(\cP}{}^{[\cM} f^{\cN]}{}_{\cQ)}{}^{\cR}$
as well. Thus ${\bf r}_{-1}$ cannot have any overlap with $({\bf adj} \times {\bf adj})_-$, the antisymmetric part of the tensor product
${\bf adj} \times {\bf adj}$, and we conclude
the embedding tensor must transform in the symmetric part,
${\bf r}_{-1}\subseteq({\bf adj} \times {\bf adj})_+$.

\section{Explicit results for $\g=\mathfrak{e}_{11-D}$} \label{explapp}

\subsection{Some useful numbers}

For each Lie algebra $\g=\frake_{11-D}$ we here give the Coxeter number (which is actually $C/2$, with $C$ given below), and two other related constants that depend on both $\g$ and the representation ${\bf r}_1={\bf s}_1={\bf t}_1$. We have
\begin{align}
(t_\alpha)_\cM{}^\cN (t^\beta)_\cN{}^\cM &= A \,\delta_\alpha{}^\beta, &
(t_\alpha)_\cM{}^\cP (t^\alpha)_\cP{}^\cN &= B \,\delta_\cM{}^\cN, & f_{\al\be\ga}f^{\al\be\de} &= -\,C\, \de_\ga{}^\de,
\end{align}
where the constants $A$, $B$ and $C$ are given by
\begin{align}
A &= \frac{\dim {\bf r}_1}{\dim \mathfrak{g}} B, &
B &= \frac{61-(D-2)^2}{D-2}, & C&=2\bigg(\frac{\dim \mathfrak{g}}{11-D}-1\bigg),
\end{align}
or alternatively, $D \neq 3$, by
\begin{align}
A &= \frac{(D-1)\,\dim {\bf r}_1}{(D-2)(11-D)}, &
B &= \frac{(D-1)\,\dim \mathfrak{g}}{(D-2)(11-D)}, &
C &= 2\bigg(\frac{D-2}{D-1}\,B-1\bigg).
\end{align}
Explicitly the values of $A$, $B$ and $C$ are given
below, together with the values of another constant $K$, which will prove useful in the next subsection (at least for $D=4,5,6$).
\setlength{\arraycolsep}{15pt}
{\renewcommand{\arraystretch}{1.5}
\begin{align*}
\begin{array}{|c|c|c|c|c|c|c|c|}
\hline
D  
& \mathfrak{g} 
& {\dim {\bf r}_1} &{\dim \mathfrak{g}}& A & B & C & K\\
\hline
7
& \mathfrak{a}_4 & 10 &24 &3& 36/5& 10 & 24\\
6
& \mathfrak{d}_5  & 16&45&4& 45/4& 16 & 20\\
5
& \mathfrak{e}_6 &27&78&6& 52/3& 24 & 20\\
4 
& \mathfrak{e}_7 &56&133&12& 57/2& 36 & 21\\
3
& \mathfrak{e}_8 &248&248&60& 60& 60 & 112/5=22.4\\
\hline
\end{array}
\end{align*}
}

\subsection{The ${\bf s}_p$ representations and some projectors} \label{app-b2}

For each representation ${\bf s}_p$ there is an associated projector $\mathbb{P}_p$.
For $D \neq 3$ we have
\begin{align} \label{2-proj}
(\mathbb{P}_2)_{\cM\cN}{}^{\cP\cQ} = \frac1{2(10-D)}\bigg(\frac{D-1}{D-2} \de_{(\cM}{}^\cP \de_{\cN)}{}^\cQ - 
(t_\al)_{(\cM}{}^{\cP}(t^\al)_{\cN)}{}^\cQ \bigg),
\end{align}
and for $D = 4,5,6$ we have
\begin{align} \label{d-1-proj}
(\mathbb{P}_{D-1})_{\cM}{}^\al{}^{\cN}{}_\be &= \frac1K\big(
(D-1)\de_\cM{}^\cN\de^\al{}_\be - (D-2)(t_\be)_\cM{}^\cP (t^\al)_\cP{}^\cN- f^\al{}_{\be\ga}(t^\ga)_\cM{}^\cN\big),
\end{align}
where the explicit value of the constant
\begin{align}
K = \frac{2(11-D)(10-D)}{8-D}
\end{align}
was given in the table in the preceding subsection.
For $D=7$ the representation ${\bf s}_{D-1}$ is reducible, ${\bf s}_{6}=\overline{\bf 40}+{\bf 15}$, and
the right hand side of (\ref{d-1-proj}) decomposes into a linear combination of the projectors of the irreducible subrepresentations,
\begin{align}
\frac32\,( \mathbb{P}_{\overline{\bf 40}})_{\cM}{}^\al{}^{\cN}{}_\be + 2 \,(\mathbb{P}_{\bf 15})_{\cM}{}^\al{}^{\cN}{}_\be.
\end{align}
Also
for $D=3$ the representation ${\bf s}_{D-1}$ is reducible, ${\bf s}_{2}={\bf 1}+{\bf 3875}$ (although ${\bf r}_{2}$ is not,
${\bf r}_{2}={\bf 3875}$). But in this case the right hand side of (\ref{d-1-proj}) contains an extra term, in addition to
the projectors of ${\bf 1}$ and ${\bf 3875}$, that mixes the two types of indices, which is possible only for $D=3$. If we instead compare $(\mathbb{P}_2)_{\cM\cN}{}^{\cP\cQ}$ for $D=3$ with the right hand side of 
(\ref{2-proj}), we find that it is equal to
\begin{align}
(\mathbb{P}_{\bf 3875})_{\cM\cN}{}^{\cP\cQ}
+ \frac{31}{7}(\mathbb{P}_{\bf 1})_{\cM\cN}{}^{\cP\cQ},
\end{align}
which is not the same as $(\mathbb{P}_2)_{\cM\cN}{}^{\cP\cQ}$, since the coefficients of the two terms are different.

Below we give explicitly the irreducible representations in the tensor product $\overline{\bf r}_1 \times {\bf adj}$ and their projectors.
For $D=7$ we split each ${\bf r}_1$ index into an antisymmetric pair of fundamental $\sl(5)$ indices.
We also give explicitly the representation ${\bf s}_2$ and its orthogonal complement
${\bf s}_\perp$ in the symmetric tensor product $({\bf r}_1 \times {\bf r}_1)_+$ for $D=4,5,6$. For $D=3$ we write the projectors of all the irreducible representations that are contained in the tensor product ${\bf 248} \times {\bf 248}$, 
which were first given in \cite{Koepsell:1999uj}. 
Finally we write all the irreducible representations contained in ${\bf s}_p$ for $3 \leq D \leq 7$, up to $p=D+1$, and the Dynkin labels of minus their lowest weights (denoted by $\Lambda$). They have been computed using the results in
\cite{Palmkvist:2012nc}, and the computer program SimpLie \cite{Bergshoeff:2007qi}, except for the $D=3$, $p=4$ representations, which were given in \cite{Kleinschmidt:2013em}.
\\\\
\noindent
$D=7:$
\begin{align}
(\tilde{\mathbb{P}}_{\bf 10})_{ab\,|\,c}{}^d\,|\,{}^{ef\,|\,g}{}_h&=\tfrac5{18}(-\de_c{}^g\de_{[a}{}^d\de_{b]}{}^{[e}\de_h{}^{f]}
-\de_{ab}{}^{gd}\de_{ch}{}^{ef})\nn\\
({\mathbb{P}}_{\bf 15})_{ab\,|\,c}{}^d\,|\,{}^{ef\,|\,g}{}_h&=\tfrac14(-\de_c{}^g\de_{[a}{}^d\de_{b]}{}^{[e}\de_h{}^{f]}
+\de_{ab}{}^{gd}\de_{ch}{}^{ef})\nn\\
({\mathbb{P}}_{\bf \overline{40}})_{ab\,|\,c}{}^d\,|\,{}^{ef\,|\,g}{}_h&=\tfrac23(\de_{abc}{}^{efg}\de_h{}^d
+2\,\de_{abch}{}^{efgd})\nn\\
(\tilde{\mathbb{P}}_{\bf 175})_{ab\,|\,c}{}^d\,|\,{}^{ef\,|\,g}{}_h&=
\tfrac23 \de_c{}^g \de_h{}^d \de_{ab}{}^{ef}-\tfrac23 \de_h{}^d\de_{[a}{}^g\de_{b]}{}^{[e}\de_c{}^{f]})\nn\\
&\quad\,+\tfrac1{36}(11  \,\de_c{}^g\de_{[a}{}^d\de_{b]}{}^{[e}\de_h{}^{f]}
-7\,\de_{ab}{}^{gd} \de_{ch}{}^{ef})
\end{align}
The tilde in the projectors of ${\bf 10}$ and ${\bf 175}$ indicates that the trace has to be taken out. For example,
\begin{align}
({\mathbb{P}}_{\bf 10})_{ab\,|\,i}{}^j\,|\,{}^{ef\,|\,k}{}_l = (\de_i{}^c \de_d{}^j -\tfrac15 \de_i{}^j\de_c{}^d)(\de_l{}^h \de_g{}^k -\tfrac15 \de_l{}^k\de_g{}^h)(\tilde{\mathbb{P}}_{\bf 10})_{ab\,|\,c}{}^d\,|\,{}^{ef\,|\,g}{}_h.
\end{align}
$D=6:$
\begin{align}
{{\bf 16}_s}&={\bf r}_{\rm S} &(\mathbb{P}_{\rm S}){}^{\cN}{}_\be{}_{\cM}{}^\al &= \tfrac4{45}(t_\be)_\cM{}^\cP (t^\al)_\cP{}^\cN
-\tfrac4{45} f^\al{}_{\be\ga}(t^\ga)_\cM{}^\cN,\nn\\
{{\bf 144}_c}&={\bf r}_{\rm M} & (\mathbb{P}_{\rm M}){}^{\cN}{}_\be{}_{\cM}{}^\al &= 
\tfrac14\de_\cM{}^\cN\de^\al{}_\be - \tfrac1{5}(t_\be)_\cM{}^\cP (t^\al)_\cP{}^\cN- \tfrac1{20} f^\al{}_{\be\ga}(t^\ga)_\cM{}^\cN,\nn\\
{{\bf 560}_s}&={\bf r}_{\rm L} &(\mathbb{P}_{\rm L}){}^{\cN}{}_\be{}_{\cM}{}^\al &= 
\tfrac34\de_\cM{}^\cN\de^\al{}_\be + \tfrac1{9}(t_\be)_\cM{}^\cP (t^\al)_\cP{}^\cN+ \tfrac{5}{36} f^\al{}_{\be\ga}(t^\ga)_\cM{}^\cN \nn
\end{align}
\begin{align}
{{\bf 10}}&={\bf s}_2&(\mathbb{P}_{2})_{\cM\cN}{}^{\cP\cQ} &= \tfrac5{32} \de_{(\cM}{}^\cP \de_{\cN)}{}^\cQ - 
\tfrac1{8}(t_\al)_{(\cM}{}^{\cP}(t^\al)_{\cN)}{}^\cQ,\nn\\
{{\bf 126}_c}&={\bf s}_\perp&(\mathbb{P}_{\perp}
)_{\cM\cN}{}^{\cP\cQ} &= \tfrac{27}{32} \de_{(\cM}{}^\cP \de_{\cN)}{}^\cQ +\tfrac1{8}
(t_\al)_{(\cM}{}^{\cP}(t^\al)_{\cN)}{}^\cQ,
\end{align}
$D=5:$
\begin{align}
{\bf 27}&={\bf r}_{\rm S} & (\mathbb{P}_{\rm S}){}^{\cN}{}_\be{}_{\cM}{}^\al &= \tfrac3{52}(t_\be)_\cM{}^\cP (t^\al)_\cP{}^\cN
-\tfrac3{52} f^\al{}_{\be\ga}(t^\ga)_\cM{}^\cN,\nn\\
{\overline{\bf 351}}&={\bf r}_{\rm M} & (\mathbb{P}_{\rm M}){}^{\cN}{}_\be{}_{\cM}{}^\al &= 
\tfrac15\de_\cM{}^\cN\de^\al{}_\be - \tfrac3{20}(t_\be)_\cM{}^\cP (t^\al)_\cP{}^\cN- \tfrac1{20} f^\al{}_{\be\ga}(t^\ga)_\cM{}^\cN,\nn\\
{\bf 1728}&={\bf r}_{\rm L} & (\mathbb{P}_{\rm L}){}^{\cN}{}_\be{}_{\cM}{}^\al &= 
\tfrac45\de_\cM{}^\cN\de^\al{}_\be + \tfrac6{65}(t_\be)_\cM{}^\cP (t^\al)_\cP{}^\cN+ \tfrac{7}{65} f^\al{}_{\be\ga}(t^\ga)_\cM{}^\cN \nn
\end{align}
\begin{align}
{\bf 27}&={\bf s}_2 & (\mathbb{P}_{2})_{\cM\cN}{}^{\cP\cQ} &= \tfrac2{15} \de_{(\cM}{}^\cP \de_{\cN)}{}^\cQ - 
\tfrac1{10}(t_\al)_{(\cM}{}^{\cP}(t^\al)_{\cN)}{}^\cQ,\nn\\
{\overline{\bf 351}'}&={\bf s}_\perp&(\mathbb{P}_{\perp})_{\cM\cN}{}^{\cP\cQ} 
&=  \tfrac{13}{15} \de_{(\cM}{}^\cP \de_{\cN)}{}^\cQ +\tfrac1{10}
(t_\al)_{(\cM}{}^{\cP}(t^\al)_{\cN)}{}^\cQ,
\end{align}
$D=4:$
\begin{align}
{\bf 56}&= {\bf r}_{\rm S} & (\mathbb{P}_{\rm S}){}^{\cN}{}_\be{}_{\cM}{}^\al &= \tfrac2{57}(t_\be)_\cM{}^\cP (t^\al)_\cP{}^\cN
-\tfrac2{57} f^\al{}_{\be\ga}(t^\ga)_\cM{}^\cN,\nn\\
{\bf 912}&={\bf r}_{\rm M} & (\mathbb{P}_{\rm M}){}^{\cN}{}_\be{}_{\cM}{}^\al &= 
\tfrac17\de_\cM{}^\cN\de^\al{}_\be - \tfrac2{21}(t_\be)_\cM{}^\cP (t^\al)_\cP{}^\cN- \tfrac1{21} f^\al{}_{\be\ga}(t^\ga)_\cM{}^\cN,\nn\\
{\bf 6480}&={\bf r}_{\rm L} &(\mathbb{P}_{\rm L}){}^{\cN}{}_\be{}_{\cM}{}^\al &= 
\tfrac67\de_\cM{}^\cN\de^\al{}_\be + \tfrac8{133}(t_\be)_\cM{}^\cP (t^\al)_\cP{}^\cN+ \tfrac{11}{133} f^\al{}_{\be\ga}(t^\ga)_\cM{}^\cN \nn
\end{align}
\begin{align}
{\bf 133}&={\bf s}_2&(\mathbb{P}_{2})_{\cM\cN}{}^{\cP\cQ} &= \tfrac18 \de_{(\cM}{}^\cP \de_{\cN)}{}^\cQ - 
\tfrac1{12}(t_\al)_{(\cM}{}^{\cP}(t^\al)_{\cN)}{}^\cQ,\nn\\
{\bf 1463}&={\bf s}_\perp&(\mathbb{P}_{\perp}
)_{\cM\cN}{}^{\cP\cQ} &= \tfrac78 \de_{(\cM}{}^\cP \de_{\cN)}{}^\cQ +\tfrac1{12}
(t_\al)_{(\cM}{}^{\cP}(t^\al)_{\cN)}{}^\cQ,
\end{align}
$D=3:$
\begin{align} \label{e8projektorer}
(\mathbb{P}_{\bf1})_{\cM\cN}{}^{\cP\cQ} &= \tfrac{1}{248}\eta_{\cM\cN}\eta^{\cP\cQ},\nn\\
(\mathbb{P}_{\bf248})_{\cM\cN}{}^{\cP\cQ} &= -\tfrac{1}{60}
f^{\cR}{}_{\cM\cN}f_{\cR}{}^{\cP\cQ},\nn\\
(\mathbb{P}_{\bf3875})_{\cM\cN}{}^{\cP\cQ} &= \tfrac{1}{7}\delta_{(\cM}{}^{\cP}\delta_{\cN)}{}^{\cQ}
-\tfrac{1}{56}\eta_{\cM\cN}\eta^{\cP\cQ}-\tfrac{1}{14}f^{\cR}{}_{\cM}{}^{(\cP}f_{\cR\cN}{}^{\cQ)},\nn\\
(\mathbb{P}_{\bf27000})_{\cM\cN}{}^{\cP\cQ} &= \tfrac{6}{7}\delta_{(\cM}{}^{\cP}\delta_{\cN)}{}^{\cQ}
+\tfrac{3}{217}\eta_{\cM\cN}\eta^{\cP\cQ}+\tfrac{1}{14}f^{\cR}{}_{\cM}{}^{(\cP}f_{\cR\cN}{}^{\cQ)},\nn\\
(\mathbb{P}_{\bf30380})_{\cM\cN}{}^{\cP\cQ} &= 
\delta_{[\cM}{}^{\cP}\delta_{\cN]}{}^{\cQ}+\tfrac{1}{60}f^{\cR}{}_{\cM\cN}f_{\cR}{}^{\cP\cQ}.
\end{align}

\newpage

\noindent
$D=7:$
\setlength{\arraycolsep}{5.97pt}
{\renewcommand{\arraystretch}{1.5}
\begin{flalign*}
\begin{array}{|c|c|c|c|c|c|cc|ccc|cccc|c}
\hline
p & 1 & 2&3 & 4&5&\multicolumn{2}{c|}{6}&\multicolumn{3}{c|}{7}&\multicolumn{4}{c|}{8}& \\
\hline
{\bf s}_p & \overline{\bf 10}&{\bf5}&\overline{\bf 5}&{\bf 10}&{\bf 24}&\overline{\bf 15}&{\bf 40}&{\bf 5}&\overline{\bf 45}&{\bf 70}
&\overline{\bf 5}&{\bf 45}&\overline{\bf 70}&{\bf 105}&\\
\hline
\Lambda& 
\begin{picture}(15,20)(0,0)
\put(0,0){${\scriptstyle 1}$}
\put(5,0){${\scriptstyle 0}$}
\put(10,0){${\scriptstyle 0}$}
\put(0,8){${\scriptstyle 0}$}
\end{picture}& 
\begin{picture}(15,20)(0,0)
\put(0,0){${\scriptstyle 0}$}
\put(5,0){${\scriptstyle 0}$}
\put(10,0){${\scriptstyle 1}$}
\put(0,8){${\scriptstyle 0}$}
\end{picture}&
\begin{picture}(15,20)(0,0)
\put(0,0){${\scriptstyle 0}$}
\put(5,0){${\scriptstyle 0}$}
\put(10,0){${\scriptstyle 0}$}
\put(0,8){${\scriptstyle 1}$}
\end{picture}&
\begin{picture}(15,20)(0,0)
\put(0,0){${\scriptstyle 0}$}
\put(5,0){${\scriptstyle 1}$}
\put(10,0){${\scriptstyle 0}$}
\put(0,8){${\scriptstyle 0}$}
\end{picture}&
\begin{picture}(15,20)(0,0)
\put(0,0){${\scriptstyle 0}$}
\put(5,0){${\scriptstyle 0}$}
\put(10,0){${\scriptstyle 1}$}
\put(0,8){${\scriptstyle 1}$}
\end{picture}&
\begin{picture}(15,20)(0,0)
\put(0,0){${\scriptstyle 0}$}
\put(5,0){${\scriptstyle 0}$}
\put(10,0){${\scriptstyle 0}$}
\put(0,8){${\scriptstyle 2}$}
\end{picture}&
\begin{picture}(15,20)(0,0)
\put(0,0){${\scriptstyle 0}$}
\put(5,0){${\scriptstyle 1}$}
\put(10,0){${\scriptstyle 1}$}
\put(0,8){${\scriptstyle 0}$}
\end{picture}&
\begin{picture}(15,20)(0,0)
\put(0,0){${\scriptstyle 0}$}
\put(5,0){${\scriptstyle 0}$}
\put(10,0){${\scriptstyle 1}$}
\put(0,8){${\scriptstyle 0}$}
\end{picture}&
\begin{picture}(15,20)(0,0)
\put(0,0){${\scriptstyle 0}$}
\put(5,0){${\scriptstyle 1}$}
\put(10,0){${\scriptstyle 0}$}
\put(0,8){${\scriptstyle 1}$}
\end{picture}&
\begin{picture}(15,20)(0,0)
\put(0,0){${\scriptstyle 0}$}
\put(5,0){${\scriptstyle 0}$}
\put(10,0){${\scriptstyle 2}$}
\put(0,8){${\scriptstyle 1}$}
\end{picture}&
\begin{picture}(15,20)(0,0)
\put(0,0){${\scriptstyle 0}$}
\put(5,0){${\scriptstyle 0}$}
\put(10,0){${\scriptstyle 0}$}
\put(0,8){${\scriptstyle 1}$}
\end{picture}&
\begin{picture}(15,20)(0,0)
\put(0,0){${\scriptstyle 1}$}
\put(5,0){${\scriptstyle 0}$}
\put(10,0){${\scriptstyle 1}$}
\put(0,8){${\scriptstyle 0}$}
\end{picture}&
\begin{picture}(15,20)(0,0)
\put(0,0){${\scriptstyle 0}$}
\put(5,0){${\scriptstyle 0}$}
\put(10,0){${\scriptstyle 1}$}
\put(0,8){${\scriptstyle 2}$}
\end{picture}&
\begin{picture}(15,20)(0,0)
\put(0,0){${\scriptstyle 0}$}
\put(5,0){${\scriptstyle 1}$}
\put(10,0){${\scriptstyle 2}$}
\put(0,8){${\scriptstyle 0}$}
\end{picture}
&\\
\hline
\end{array} &&
\end{flalign*}
}

\noindent
$D=6:$
\setlength{\arraycolsep}{5.12pt}
{\renewcommand{\arraystretch}{1.5}
\begin{flalign*}
\begin{array}{|c|c|c|c|c|c|ccc|cccc|c}
\hline
p &1 & 2&3 & 4&5&\multicolumn{3}{c|}{6}&\multicolumn{4}{c|}{7}&
\\
\hline
{\bf s}_p & {\bf 16}_c&{\bf10}&{\bf 16}_s&{\bf45}&{\bf 144}_s&{\bf 10}&{\bf 126}_s&{\bf 320}&{\bf 16}_s & {\bf 144}_c&{\bf 560}_s&{\bf 720}_s&\\
\hline
\Lambda&
\begin{picture}(20,20)(0,0)
\put(0,0){${\scriptstyle 1}$}
\put(5,0){${\scriptstyle 0}$}
\put(10,0){${\scriptstyle 0}$}
\put(15,0){${\scriptstyle 0}$}
\put(5,8){${\scriptstyle 0}$}
\end{picture}& 
\begin{picture}(20,20)(0,0)
\put(0,0){${\scriptstyle 0}$}
\put(5,0){${\scriptstyle 0}$}
\put(10,0){${\scriptstyle 0}$}
\put(15,0){${\scriptstyle 1}$}
\put(5,8){${\scriptstyle 0}$}
\end{picture}& 
\begin{picture}(20,20)(0,0)
\put(0,0){${\scriptstyle 0}$}
\put(5,0){${\scriptstyle 0}$}
\put(10,0){${\scriptstyle 0}$}
\put(15,0){${\scriptstyle 0}$}
\put(5,8){${\scriptstyle 1}$}
\end{picture}& 
\begin{picture}(20,20)(0,0)
\put(0,0){${\scriptstyle 0}$}
\put(5,0){${\scriptstyle 0}$}
\put(10,0){${\scriptstyle 1}$}
\put(15,0){${\scriptstyle 0}$}
\put(5,8){${\scriptstyle 0}$}
\end{picture}& 
\begin{picture}(20,20)(0,0)
\put(0,0){${\scriptstyle 0}$}
\put(5,0){${\scriptstyle 0}$}
\put(10,0){${\scriptstyle 0}$}
\put(15,0){${\scriptstyle 1}$}
\put(5,8){${\scriptstyle 1}$}
\end{picture}&
\begin{picture}(20,20)(0,0)
\put(0,0){${\scriptstyle 0}$}
\put(5,0){${\scriptstyle 0}$}
\put(10,0){${\scriptstyle 0}$}
\put(15,0){${\scriptstyle 1}$}
\put(5,8){${\scriptstyle 0}$}
\end{picture}&
\begin{picture}(20,20)(0,0)
\put(0,0){${\scriptstyle 0}$}
\put(5,0){${\scriptstyle 0}$}
\put(10,0){${\scriptstyle 0}$}
\put(15,0){${\scriptstyle 0}$}
\put(5,8){${\scriptstyle 2}$}
\end{picture}&
\begin{picture}(20,20)(0,0)
\put(0,0){${\scriptstyle 0}$}
\put(5,0){${\scriptstyle 0}$}
\put(10,0){${\scriptstyle 1}$}
\put(15,0){${\scriptstyle 1}$}
\put(5,8){${\scriptstyle 0}$}
\end{picture}&
\begin{picture}(20,20)(0,0)
\put(0,0){${\scriptstyle 0}$}
\put(5,0){${\scriptstyle 0}$}
\put(10,0){${\scriptstyle 0}$}
\put(15,0){${\scriptstyle 0}$}
\put(5,8){${\scriptstyle 1}$}
\end{picture}&
\begin{picture}(20,20)(0,0)
\put(0,0){${\scriptstyle 1}$}
\put(5,0){${\scriptstyle 0}$}
\put(10,0){${\scriptstyle 0}$}
\put(15,0){${\scriptstyle 1}$}
\put(5,8){${\scriptstyle 0}$}
\end{picture}&
\begin{picture}(20,20)(0,0)
\put(0,0){${\scriptstyle 0}$}
\put(5,0){${\scriptstyle 0}$}
\put(10,0){${\scriptstyle 1}$}
\put(15,0){${\scriptstyle 0}$}
\put(5,8){${\scriptstyle 1}$}
\end{picture}&
\begin{picture}(20,20)(0,0)
\put(0,0){${\scriptstyle 0}$}
\put(5,0){${\scriptstyle 0}$}
\put(10,0){${\scriptstyle 0}$}
\put(15,0){${\scriptstyle 2}$}
\put(5,8){${\scriptstyle 1}$}
\end{picture}&\\
\hline
\end{array} &&
\end{flalign*}
}

\noindent
$D=5:$
\setlength{\arraycolsep}{4.85pt}
{\renewcommand{\arraystretch}{1.5}
\begin{flalign*}
\begin{array}{|c|c|c|c|c|cc|ccccc|c}
\hline
p & 1 & 2&3 &4& \multicolumn{2}{c|}{5}&\multicolumn{5}{c|}{6}&\\
\hline
{\bf s}_p &\overline{\bf 27}&{\bf27}&{\bf 78}&{\bf351}&{\bf 27}&{\bf 1728}&{\bf 1}&{\bf 78}&{\bf 650}&{\bf 2430}&{\bf 5824}&\\
\hline
\Lambda&
\begin{picture}(25,20)(0,0)
\put(0,0){${\scriptstyle 1}$}
\put(5,0){${\scriptstyle 0}$}
\put(10,0){${\scriptstyle 0}$}
\put(15,0){${\scriptstyle 0}$}
\put(20,0){${\scriptstyle 0}$}
\put(10,8){${\scriptstyle 0}$}
\end{picture}& 
\begin{picture}(25,20)(0,0)
\put(0,0){${\scriptstyle 0}$}
\put(5,0){${\scriptstyle 0}$}
\put(10,0){${\scriptstyle 0}$}
\put(15,0){${\scriptstyle 0}$}
\put(20,0){${\scriptstyle 1}$}
\put(10,8){${\scriptstyle 0}$}
\end{picture}& 
\begin{picture}(25,20)(0,0)
\put(0,0){${\scriptstyle 0}$}
\put(5,0){${\scriptstyle 0}$}
\put(10,0){${\scriptstyle 0}$}
\put(15,0){${\scriptstyle 0}$}
\put(20,0){${\scriptstyle 0}$}
\put(10,8){${\scriptstyle 1}$}
\end{picture}& 
\begin{picture}(25,20)(0,0)
\put(0,0){${\scriptstyle 0}$}
\put(5,0){${\scriptstyle 0}$}
\put(10,0){${\scriptstyle 0}$}
\put(15,0){${\scriptstyle 1}$}
\put(20,0){${\scriptstyle 0}$}
\put(10,8){${\scriptstyle 0}$}
\end{picture}& 
\begin{picture}(25,20)(0,0)
\put(0,0){${\scriptstyle 0}$}
\put(5,0){${\scriptstyle 0}$}
\put(10,0){${\scriptstyle 0}$}
\put(15,0){${\scriptstyle 0}$}
\put(20,0){${\scriptstyle 1}$}
\put(10,8){${\scriptstyle 0}$}
\end{picture}&
\begin{picture}(25,20)(0,0)
\put(0,0){${\scriptstyle 0}$}
\put(5,0){${\scriptstyle 0}$}
\put(10,0){${\scriptstyle 0}$}
\put(15,0){${\scriptstyle 0}$}
\put(20,0){${\scriptstyle 1}$}
\put(10,8){${\scriptstyle 1}$}
\end{picture}&\begin{picture}(25,20)(0,0)
\put(0,0){${\scriptstyle 0}$}
\put(5,0){${\scriptstyle 0}$}
\put(10,0){${\scriptstyle 0}$}
\put(15,0){${\scriptstyle 0}$}
\put(20,0){${\scriptstyle 0}$}
\put(10,8){${\scriptstyle 0}$}
\end{picture}& 
\begin{picture}(25,20)(0,0)
\put(0,0){${\scriptstyle 0}$}
\put(5,0){${\scriptstyle 0}$}
\put(10,0){${\scriptstyle 0}$}
\put(15,0){${\scriptstyle 0}$}
\put(20,0){${\scriptstyle 0}$}
\put(10,8){${\scriptstyle 1}$}
\end{picture}& 
\begin{picture}(25,20)(0,0)
\put(0,0){${\scriptstyle 1}$}
\put(5,0){${\scriptstyle 0}$}
\put(10,0){${\scriptstyle 0}$}
\put(15,0){${\scriptstyle 0}$}
\put(20,0){${\scriptstyle 1}$}
\put(10,8){${\scriptstyle 0}$}
\end{picture}& 
\begin{picture}(25,20)(0,0)
\put(0,0){${\scriptstyle 0}$}
\put(5,0){${\scriptstyle 0}$}
\put(10,0){${\scriptstyle 0}$}
\put(15,0){${\scriptstyle 0}$}
\put(20,0){${\scriptstyle 0}$}
\put(10,8){${\scriptstyle 2}$}
\end{picture}&
\begin{picture}(25,20)(0,0)
\put(0,0){${\scriptstyle 0}$}
\put(5,0){${\scriptstyle 0}$}
\put(10,0){${\scriptstyle 0}$}
\put(15,0){${\scriptstyle 1}$}
\put(20,0){${\scriptstyle 1}$}
\put(10,8){${\scriptstyle 0}$}
\end{picture}&
\\
\hline
\end{array} &&
\end{flalign*}
}

\noindent
$D=4:$
\setlength{\arraycolsep}{6.08pt}
{\renewcommand{\arraystretch}{1.5}
\begin{flalign*}
\begin{array}{|c|c|c|c|cc|cccc|c}
\hline
p & 1 & 2&3 & \multicolumn{2}{c|}{4}&\multicolumn{4}{c|}{5}&
\\
\hline
{\bf s}_p & {\bf56}&{\bf133}&{\bf912}&{\bf133}&{\bf8645}&{\bf 56}&{\bf 912}&{\bf 6480}&{\bf 86184}&\\
\hline
\Lambda&
\begin{picture}(30,20)(0,0)
\put(0,0){${\scriptstyle 1}$}
\put(5,0){${\scriptstyle 0}$}
\put(10,0){${\scriptstyle 0}$}
\put(15,0){${\scriptstyle 0}$}
\put(20,0){${\scriptstyle 0}$}
\put(25,0){${\scriptstyle 0}$}
\put(15,8){${\scriptstyle 0}$}
\end{picture}& 
\begin{picture}(30,20)(0,0)
\put(0,0){${\scriptstyle 0}$}
\put(5,0){${\scriptstyle 0}$}
\put(10,0){${\scriptstyle 0}$}
\put(15,0){${\scriptstyle 0}$}
\put(20,0){${\scriptstyle 0}$}
\put(25,0){${\scriptstyle 1}$}
\put(15,8){${\scriptstyle 0}$}
\end{picture}& 
\begin{picture}(30,20)(0,0)
\put(0,0){${\scriptstyle 0}$}
\put(5,0){${\scriptstyle 0}$}
\put(10,0){${\scriptstyle 0}$}
\put(15,0){${\scriptstyle 0}$}
\put(20,0){${\scriptstyle 0}$}
\put(25,0){${\scriptstyle 0}$}
\put(15,8){${\scriptstyle 1}$}
\end{picture}& 
\begin{picture}(30,20)(0,0)
\put(0,0){${\scriptstyle 0}$}
\put(5,0){${\scriptstyle 0}$}
\put(10,0){${\scriptstyle 0}$}
\put(15,0){${\scriptstyle 0}$}
\put(20,0){${\scriptstyle 0}$}
\put(25,0){${\scriptstyle 1}$}
\put(15,8){${\scriptstyle 0}$}
\end{picture}& 
\begin{picture}(30,20)(0,0)
\put(0,0){${\scriptstyle 0}$}
\put(5,0){${\scriptstyle 0}$}
\put(10,0){${\scriptstyle 0}$}
\put(15,0){${\scriptstyle 0}$}
\put(20,0){${\scriptstyle 1}$}
\put(25,0){${\scriptstyle 0}$}
\put(15,8){${\scriptstyle 0}$}
\end{picture}&
\begin{picture}(30,20)(0,0)
\put(0,0){${\scriptstyle 1}$}
\put(5,0){${\scriptstyle 0}$}
\put(10,0){${\scriptstyle 0}$}
\put(15,0){${\scriptstyle 0}$}
\put(20,0){${\scriptstyle 0}$}
\put(25,0){${\scriptstyle 0}$}
\put(15,8){${\scriptstyle 0}$}
\end{picture}& 
\begin{picture}(30,20)(0,0)
\put(0,0){${\scriptstyle 0}$}
\put(5,0){${\scriptstyle 0}$}
\put(10,0){${\scriptstyle 0}$}
\put(15,0){${\scriptstyle 0}$}
\put(20,0){${\scriptstyle 0}$}
\put(25,0){${\scriptstyle 0}$}
\put(15,8){${\scriptstyle 1}$}
\end{picture}& 
\begin{picture}(30,20)(0,0)
\put(0,0){${\scriptstyle 1}$}
\put(5,0){${\scriptstyle 0}$}
\put(10,0){${\scriptstyle 0}$}
\put(15,0){${\scriptstyle 0}$}
\put(20,0){${\scriptstyle 0}$}
\put(25,0){${\scriptstyle 1}$}
\put(15,8){${\scriptstyle 0}$}
\end{picture}& 
\begin{picture}(30,20)(0,0)
\put(0,0){${\scriptstyle 0}$}
\put(5,0){${\scriptstyle 0}$}
\put(10,0){${\scriptstyle 0}$}
\put(15,0){${\scriptstyle 0}$}
\put(20,0){${\scriptstyle 0}$}
\put(25,0){${\scriptstyle 1}$}
\put(15,8){${\scriptstyle 1}$}
\end{picture}&\\
\hline
\end{array} &&
\end{flalign*}
}

\noindent
$D=3:$
\setlength{\arraycolsep}{4.85pt}
{\renewcommand{\arraystretch}{1.5}
\begin{flalign*}
\begin{array}{|c|c|cc|ccc|c}
\hline
p &1 & \multicolumn{2}{c|}{2} & \multicolumn{3}{c|}{3}&\qquad\qquad \qquad 4\quad  \qquad\quad\!\!\!\!\!\\
\hline
{\bf s}_p & {\bf248}&{\bf1}&{\bf3875}&{\bf248}&{\bf3875}&{\bf147250}&\multirow{2}{*}{\it $\ $ see below}\\
\cline{1-7}
\Lambda&
\begin{picture}(35,20)(0,0)
\put(0,0){${\scriptstyle 1}$}
\put(5,0){${\scriptstyle 0}$}
\put(10,0){${\scriptstyle 0}$}
\put(15,0){${\scriptstyle 0}$}
\put(20,0){${\scriptstyle 0}$}
\put(25,0){${\scriptstyle 0}$}
\put(30,0){${\scriptstyle 0}$}
\put(20,7){${\scriptstyle 0}$}
\end{picture}& 
\begin{picture}(35,20)(0,0)
\put(0,0){${\scriptstyle 0}$}
\put(5,0){${\scriptstyle 0}$}
\put(10,0){${\scriptstyle 0}$}
\put(15,0){${\scriptstyle 0}$}
\put(20,0){${\scriptstyle 0}$}
\put(25,0){${\scriptstyle 0}$}
\put(30,0){${\scriptstyle 0}$}
\put(20,7){${\scriptstyle 0}$}
\end{picture}& 
\begin{picture}(35,20)(0,0)
\put(0,0){${\scriptstyle 0}$}
\put(5,0){${\scriptstyle 0}$}
\put(10,0){${\scriptstyle 0}$}
\put(15,0){${\scriptstyle 0}$}
\put(20,0){${\scriptstyle 0}$}
\put(25,0){${\scriptstyle 0}$}
\put(30,0){${\scriptstyle 1}$}
\put(20,7){${\scriptstyle 0}$}
\end{picture}& 
\begin{picture}(35,20)(0,0)
\put(0,0){${\scriptstyle 1}$}
\put(5,0){${\scriptstyle 0}$}
\put(10,0){${\scriptstyle 0}$}
\put(15,0){${\scriptstyle 0}$}
\put(20,0){${\scriptstyle 0}$}
\put(25,0){${\scriptstyle 0}$}
\put(30,0){${\scriptstyle 0}$}
\put(20,7){${\scriptstyle 0}$}
\end{picture}& 
\begin{picture}(35,20)(0,0)
\put(0,0){${\scriptstyle 0}$}
\put(5,0){${\scriptstyle 0}$}
\put(10,0){${\scriptstyle 0}$}
\put(15,0){${\scriptstyle 0}$}
\put(20,0){${\scriptstyle 0}$}
\put(25,0){${\scriptstyle 0}$}
\put(30,0){${\scriptstyle 1}$}
\put(20,7){${\scriptstyle 0}$}
\end{picture}& 
\begin{picture}(35,20)(0,0)
\put(0,0){${\scriptstyle 0}$}
\put(5,0){${\scriptstyle 0}$}
\put(10,0){${\scriptstyle 0}$}
\put(15,0){${\scriptstyle 0}$}
\put(20,0){${\scriptstyle 0}$}
\put(25,0){${\scriptstyle 0}$}
\put(30,0){${\scriptstyle 0}$}
\put(20,7){${\scriptstyle 1}$}
\end{picture}
&\\
\hline
\end{array} &&
\end{flalign*}
}

\noindent
$D=3$ {\it (continued)} : 
\setlength{\arraycolsep}{9.42pt}
{\renewcommand{\arraystretch}{1.5}
\begin{flalign*}
\begin{array}{|c|cccccc|c}
\hline
p & \multicolumn{6}{c|}{4}& \\
\hline
{\bf s}_p & 2\times {\bf 248}&{\bf 3875}&2\times {\bf 30380}&{\bf 147250}&{\bf 779247}&{\bf 6696000}& \\
\hline
\Lambda&
\begin{picture}(35,20)(0,0)
\put(0,0){${\scriptstyle 1}$}
\put(5,0){${\scriptstyle 0}$}
\put(10,0){${\scriptstyle 0}$}
\put(15,0){${\scriptstyle 0}$}
\put(20,0){${\scriptstyle 0}$}
\put(25,0){${\scriptstyle 0}$}
\put(30,0){${\scriptstyle 0}$}
\put(20,7){${\scriptstyle 0}$}
\end{picture}& 
\begin{picture}(35,20)(0,0)
\put(0,0){${\scriptstyle 0}$}
\put(5,0){${\scriptstyle 0}$}
\put(10,0){${\scriptstyle 0}$}
\put(15,0){${\scriptstyle 0}$}
\put(20,0){${\scriptstyle 0}$}
\put(25,0){${\scriptstyle 0}$}
\put(30,0){${\scriptstyle 1}$}
\put(20,7){${\scriptstyle 0}$}
\end{picture}& 
\begin{picture}(35,20)(0,0)
\put(0,0){${\scriptstyle 0}$}
\put(5,0){${\scriptstyle 1}$}
\put(10,0){${\scriptstyle 0}$}
\put(15,0){${\scriptstyle 0}$}
\put(20,0){${\scriptstyle 0}$}
\put(25,0){${\scriptstyle 0}$}
\put(30,0){${\scriptstyle 0}$}
\put(20,7){${\scriptstyle 0}$}
\end{picture}& 
\begin{picture}(35,20)(0,0)
\put(0,0){${\scriptstyle 0}$}
\put(5,0){${\scriptstyle 0}$}
\put(10,0){${\scriptstyle 0}$}
\put(15,0){${\scriptstyle 0}$}
\put(20,0){${\scriptstyle 0}$}
\put(25,0){${\scriptstyle 0}$}
\put(30,0){${\scriptstyle 0}$}
\put(20,7){${\scriptstyle 1}$}
\end{picture}& 
\begin{picture}(35,20)(0,0)
\put(0,0){${\scriptstyle 1}$}
\put(5,0){${\scriptstyle 0}$}
\put(10,0){${\scriptstyle 0}$}
\put(15,0){${\scriptstyle 0}$}
\put(20,0){${\scriptstyle 0}$}
\put(25,0){${\scriptstyle 0}$}
\put(30,0){${\scriptstyle 1}$}
\put(20,7){${\scriptstyle 0}$}
\end{picture}& 
\begin{picture}(35,20)(0,0)
\put(0,0){${\scriptstyle 0}$}
\put(5,0){${\scriptstyle 0}$}
\put(10,0){${\scriptstyle 0}$}
\put(15,0){${\scriptstyle 0}$}
\put(20,0){${\scriptstyle 0}$}
\put(25,0){${\scriptstyle 1}$}
\put(30,0){${\scriptstyle 0}$}
\put(20,7){${\scriptstyle 0}$}
\end{picture}
&\\
\hline
\end{array} &&
\end{flalign*}
}

\bibliographystyle{utphysmod2}

  


\providecommand{\href}[2]{#2}\begingroup\raggedright\endgroup

\end{document}